\documentclass[12pt]{article}
\usepackage{amssymb}
\usepackage{amsmath}
\usepackage{epsf}
\usepackage{epsfig}
\usepackage{cite}
\usepackage[latin1]{inputenc}
\newcommand{\lsim}{
\mathrel{\hbox{\rlap{\hbox{\lower4pt\hbox{$\sim$}}}\hbox{$<$}}}}

\newcommand{\gsim}{
\mathrel{\hbox{\rlap{\hbox{\lower4pt\hbox{$\sim$}}}\hbox{$>$}}}}

\def\Adel{\mathcal{A}_{\Delta\Gamma}}

\def\D0{D\O }
\def\mixAng{\varphi_{\rm M}}

\begin{document}
\begin{titlepage}
\vspace*{-1.3truecm}

\begin{flushright}
Nikhef-2011-024 \\
DSF/9/2011
\end{flushright}

\vspace*{1.8truecm}

\begin{center}
\boldmath
{\Large{\bf Anatomy of $B^0_{s,d} \to J/\psi f_0(980)$}}
\unboldmath
\end{center}

\vspace{0.9truecm}

\begin{center}
{\bf Robert Fleischer,\,${}^a$ Robert Knegjens\,${}^a$ \,and\, Giulia Ricciardi\,${}^{b}$}

\vspace{0.5truecm}

${}^a${\sl Nikhef, Science Park 105,
NL-1098 XG Amsterdam, The Netherlands}

\vspace{0.3truecm}

${}^b${\sl Dipartimento di Scienze Fisiche, Universit\`a di Napoli Federico II  and
I.N.F.N., Sezione di Napoli,
Complesso Universitario di Monte Sant'Angelo, Via Cintia, I-80126~Napoli, Italy}

\end{center}

\vspace{1.6cm}
\begin{abstract}
\vspace{0.2cm}\noindent
The $B^0_s\to J/\psi f_0(980)$ decay offers an interesting experimental alternative to the
well-known $B^0_s\to J/\psi \phi$ channel for the search of CP-violating New-Physics
contributions to $B^0_s$--$\bar B^0_s$ mixing. As the hadronic structure of the $f_0(980)$
has not yet been settled, we take a critical look at the implications for the relevant observables
and address recent experimental data. It turns out that the effective lifetime of
$B^0_s\to J/\psi f_0(980)$ and its mixing-induced CP asymmetry $S$ are quite robust with
respect to hadronic effects and thereby allow us to search for a large CP-violating
$B^0_s$--$\bar B^0_s$ mixing phase $\phi_s$, which is tiny in the Standard Model.
However, should small CP violation, i.e.\ in the range $-0.1\lsim S\lsim 0$, be found in
$B^0_s\to J/\psi f_0(980)$, it will be crucial to constrain hadronic corrections in order to
distinguish possible New-Physics effects from the Standard Model. We point out that
$B^0_d\to J/\psi f_0(980)$, which has not yet been measured, is a key channel in this
respect and discuss the physics potential of this decay.
\end{abstract}

\vspace*{0.5truecm}
\vfill
\noindent
September 2011
\vspace*{0.5truecm}

\end{titlepage}

\thispagestyle{empty}
\vbox{}
\newpage

\setcounter{page}{1}

\section{Introduction}\label{sec:intro}
\setcounter{equation}{0}
With the Large Hadron Collider (LHC) at CERN now collecting copious amounts of data,
the testing of the Standard Model (SM) has entered a new phase. Concerning the
quark-flavour sector, the decay $B^0_s\to J/\psi f_0(980)$, which we abbreviate from here on as
$B^0_s\to J/\psi f_0$, offers an interesting probe of CP violation  \cite{SZ}.
In particular, this channel allows us to search for
CP-violating New-Physics (NP) contributions to $B^0_s$--$\bar B^0_s$ mixing,
which is conventionally studied via the $B^0_s\to J/\psi \phi$ decay.
The draw back of the $B^0_s\to J/\psi \phi$ mode is that its final state contains two
vector mesons and is thereby a mixture of CP-even and CP-odd eigenstates.
Consequently, in order to disentangle the CP eigenstates, a time-dependent angular analysis of the
decay products $J/\psi \to \mu^+\mu^-$ and $\phi\to K^+K^-$ is necessary  \cite{DDF,DFN}.
In contrast, because the $f_0(980)$ is a scalar state with quantum numbers $J^{PC}=0^{++}$
\cite{PDG}, the final state of $B^0_s\to J/\psi f_0$ is a $p$-wave state with the CP eigenvalue
$-1$ and thus an angular analysis is not needed \cite{SZ}.

The $B^0_s\to J/\psi f_0$ channel was observed by the LHCb and Belle experiments
in the spring of 2011 \cite{LHCb-f0,Belle-f0}. These results have recently been confirmed
by the \D0 \cite{D0} (with a preliminary measurement) and CDF
collaborations \cite{CDF-f0}.
In Table~\ref{tab:BRs}, we list the reported data for $B^0_s\to J/\psi f_0$ with
$f_0\to\pi^+\pi^-$, which is the dominant channel.
LHCb, \D0 and CDF do not measure the branching ratio directly, but instead its fraction, $R_
{f_0/\phi}$, with respect to the branching ratio for $B^0_s\to J/\psi \phi$ with $\phi\to K^+ K^-$.
For comparison, we have also included in the table the corresponding branching ratios of these
fractions using additional experimental input from Ref.~\cite{PDG}.

We observe that the number of events for
$B^0_s\to J/\psi f_0$ with $f_0\to \pi^+\pi^-$ is about four times smaller than for
$B^0_s\to J/\psi \phi$ with $\phi\to K^+K^-$. Nevertheless, as no angular analysis is required,
the $B^0_s\to J/\psi f_0$ channel offers a convenient alternative to the conventional
$B^0_s\to J/\psi \phi$ decay from an experimental point of view \cite{SZ}. In
addition to a branching ratio
result, the CDF collaboration has reported a first measurement for the effective
$B^0_s\to J/\psi f_0$ lifetime \cite{CDF-f0}, and the LHCb collaboration has very recently
presented a first preliminary analysis of CP violation in $B^0_s\to J/\psi f_0$ \cite{LHCb-f0-CP}.
In the future, we should see more precise measurements of the corresponding observables.

\begin{table}[b]
\newcommand\T{\rule{0pt}{2.6ex}} 
\centering
\begin{tabular}{lcc}
\hline\hline
Experiment \T & $R_{f_0/\phi}$ & ${\rm BR}(B_s^0\to J/\psi f_0; f_0\to \pi^+ \pi^-)$ $[10^{-4}]$ \\[2pt]
\hline
LHCb \cite{LHCb-f0} \T & $0.252^{+0.046}_{-0.032}{}^{+0.027}_{-0.033}$ &
$1.68^{+0.71}_{-0.69}{}^\star$  \\[2pt]
Belle \cite{Belle-f0} & & $1.16^{+0.31}_{-0.19}{}^{+0.30}_{-0.25}$  \\[2pt]
\D0 \cite{D0} & $0.210 \pm 0.032\pm 0.036$ & $1.40\pm 0.61{}^\star$  \\[2pt]
CDF \cite{CDF-f0} & $0.257 \pm 0.020\pm0.014$ & $1.71\pm 0.65{}^\star$   \\[2pt]
\hline
\end{tabular}
\caption{Compilation of branching ratio measurements involving $B_s^0 \to J/\psi f_0$. Here
$R_{f_0/\phi} \equiv {\rm BR}(B_s^0\to J/\psi f_0; f_0\to \pi^+ \pi^-)/
{\rm BR}(B_s^0\to J/\psi \phi; \phi\to K^+ K^-)$, and a $\star$ indicates
that this result was calculated by us, for comparison, using the additional inputs
${\rm BR}(B^0_s\to J/\psi\phi)=(1.4\pm0.5) \times10^{-3}$ and
${\rm BR}(\phi\to K^+ K^-)=(48.9\pm0.5) \times 10^{-2}$ \cite{PDG}. The reported errors
are either the statistical and systematic uncertainties, respectively, or everything combined
in quadrature.}
\label{tab:BRs}
\end{table}

In view of these promising developments, we briefly summarize the current knowledge about
the $f_0(980)$ in Section~\ref{sec:f0}, and have a closer look at the $B^0_s\to J/\psi f_0$
amplitude structure in Section~\ref{sec:ampl}. In Section~\ref{sec:lifetime}, we discuss
the effective $B^0_s\to J/\psi f_0$ lifetime $\tau_{J/\psi f_0}$, which can be determined
from untagged $B_s$ data samples, derive a general bound on $\tau_{J/\psi f_0}$
and show the dependence on the CP-violating $B^0_s$--$\bar B^0_s$ mixing phase $\phi_s$.
The mixing phase also plays a key role for the time-dependent CP asymmetry of
$B^0_s\to J/\psi f_0$, which we address in Section~\ref{sec:CP}. The theoretical predictions given
in Sections~\ref{sec:lifetime} and \ref{sec:CP} are limited by doubly Cabibbo-suppressed hadronic
contributions. In Section~\ref{sec:Bd}, we point out that these effects can be constrained by
an analysis of the $B^0_d\to J/\psi f_0(980)$ channel, which has not yet been observed
and would be an interesting addition to the experimental agenda. Finally, we summarize our
results in Section~\ref{sec:concl}.

\boldmath
\section{Hadronic Structure of the $f_0(980)$}\label{sec:f0}
\unboldmath
\setcounter{equation}{0}
\subsection{Preliminaries}
Contrary to the continuing search for an elementary scalar particle, a variety of scalar hadronic
bound states have long been observed.
These states are often categorized according to whether their mass falls above or
below 1 GeV. Those belonging to the former category are expected to be composed
predominantly of quark--antiquark states and among them a $SU(3)_{\rm F}$ flavour nonet can
be identified. Unfortunately, the $f_0(980)$ belongs to the latter category, where, as we will see,
the interpretation is far from being straightforward.
The $f_0(980)$ is an isospin singlet with a mass of $(980 \pm 10)$ MeV, just  below the
$K \bar K$ threshold, and a full width between 40 MeV and 100 MeV, which reflects the fact that the
width determination is very model-dependent \cite{PDG}.

In the literature, the hadronic structure of the $f_0(980)$ state has been discussed for
decades and there are may different interpretations, from the conventional
quark--antiquark picture \cite{torn} to multiquark  \cite{Jaffe:1976ig, Hooft:2008we} or
$K\bar K$ bound states \cite{Weinstein:1982gc} (for a review, see, for instance,
Ref.~\cite{PDG} and references within). As the goal is to use the $B^0_s\to J/\psi f_0$
decay for precision tests of the CP-violating sector of the SM, it is a natural and important question
to explore how the hadronic structure of the $f_0(980)$ affects the corresponding observables.
In this section, we have a closer look at popular descriptions of the $f_0(980)$, setting
the stage for the discussion of the $B^0_s\to J/\psi f_0$ observables. We will focus on
two specific frameworks: the quark--antiquark and tetraquark pictures.

\boldmath
\subsection{The $f_0(980)$ as a Quark--Antiquark State}\label{sec:qqbar}
\unboldmath
In the conventional quark model, the scalar hadronic states are interpreted as mesons,
i.e.\ quark--antiquark ($q\bar q$) bound states, with an orbital angular momentum of $L=1$
and a spin of $S=1$ coupled to give a total angular momentum of $J=0$.
In analogy to the pseudo-scalar mesons, it is suggestive to group the observed scalar states into
nonets of the $SU(3)_{\rm F}$ flavour symmetry of strong interactions.

For the scalar states with masses $\lsim 1$~GeV, we can identify
an isotriplet $a_0(980)$, two strange isodoublets,
$\kappa$ or $K_0^\star(800)$, and two isosinglets, $\sigma$ or $f_0(600)$ and  $ f_0(980)$.
In the na\"ive quark model, assuming ideal mixing between the heaviest and lightest members
of the $SU(3)_{\rm F}$ nonet, the $f_0(980)$ and $\sigma(600)$, respectively, their quark-flavour
composition would simply be given by
\begin{equation}
|f_0(980)\rangle= |s\bar s\rangle, \quad |\sigma(600)\rangle=|n\bar n\rangle,
\end{equation}
where
\begin{equation}\label{nnbar}
|n \bar n\rangle \equiv \frac{1}{\sqrt{2}}\left(|u\bar u\rangle + |d \bar d\rangle \right)
\end{equation}
is the isospin-singlet combination of the $u\bar u$ and $d\bar d$ components.
However, there is also experimental evidence for a non-strange component of the $f_0(980)$,
which could be interpreted as evidence for the following mixing structure:
\begin{equation}\label{f0-structure}
\left(
\begin{array}{c}
|f_0 (980) \rangle\\
|\sigma (600) \rangle
\end{array}\right)
=
\left(
\begin{array}{cc}
\cos \mixAng\ &  \ \sin \mixAng \\
 -\sin \mixAng\ & \ \cos \mixAng
\end{array}
\right)
\cdot
\left(
\begin{array}{c}
|s  \bar s  \rangle \\
| n \bar n  \rangle
\end{array}
\right).
\end{equation}
Here the mixing angle $\mixAng$ is the counterpart of the $\eta$--$\eta^\prime$ mixing angle
in the standard pseudo-scalar nonet.\footnote{For a recent review on the $\eta$--$\eta^\prime$
mixing angle, see, for example,  Ref.~\cite{DiDonato:2011kr}.}

The determination of $\mixAng$ is  affected by large errors and appears process and
model dependent. Let us give  a few examples:
\begin{itemize}
\item Using $D_s^+ \to \pi^+ \pi^+ \pi^- $ transitions caused dominantly by
$D_s^+ \to \pi^+ \bar s s $ processes, the range $35^\circ \leq  |\mixAng |\leq 55^\circ$ was
estimated in Ref.~\cite{Anisovich:2003up}.
\item By making a simultaneous calculation of radiative decays of the kind
$f_0(980) \to \gamma \gamma $ and $\phi (1020) \to \gamma f_0(980)$,
$\mixAng = (4\pm 3)^\circ$ or $\mixAng = (138 \pm 6)^\circ$ were obtained
in Ref.~\cite{Anisovich:2001zp}.
\item In Ref.~\cite{Delbourgo:1998gi}, it was found that a value of
$\mixAng \simeq 20^\circ$ is consistent with the resonance data from
$\phi(1020) \to \gamma \pi^0 \pi^0 $ and  $ J/\psi \to \omega \pi \pi$ decays.
\item Using two different methods to fit the $D_{(s)} \to f_0(980)\{\pi,K\}$  branching ratios
(covariant light-front dynamics and dispersion relations),
$\mixAng = (31.5 \pm 5.0)^\circ$ and $\mixAng =(41.6 \pm 7.1)^\circ$
were obtained in Ref.~\cite{ElBennich:2008xy}.
\end{itemize}
Despite the unsatisfactory picture for the mixing angle, these studies show that the
$f_0(980)$ has a significant $s\bar s$ component. This feature is also supported
by the recent observation of the $B^0_s\to J/\psi f_0$ channel, with measurements as
summarized in Table~\ref{tab:BRs}.

Due to their non-zero orbital angular momentum, the scalar mesons are expected to
be heavier than the pseudo-scalar and vector mesons in the na\"ive quark picture.
This is not, however, the case for the light scalars that have masses below $1$~GeV.
Furthermore, the light scalar mass spectrum bears little resemblance with
that of a standard nonet.
An attractive framework to overcome these phenomenological problems is offered by 
the tetraquark model.

\boldmath
\subsection{The $f_0(980)$ as a Tetraquark}\label{sec:tetra}
\unboldmath
In the tetraquark picture, scalar states with quantum numbers
$J^{PC}=0^{++}$ are formed by the binding of diquark and anti-diquark
configurations. A diquark, denoted by $[qq']$, transforms as $\mbox{\boldmath$\bar 3$}$
under $SU(3)_{\rm C}$ colour symmetry, has
spin $S=0$, and transforms as $\mbox{\boldmath$\bar 3$}$ under $SU(3)_{\rm F}$ flavour
symmetry. Anti-diquarks, denoted by $[\bar q\bar q']$,
are in the corresponding conjugate representations. The bound scalar states of diquarks
and anti-diquarks, which do not require a non-vanishing angular momentum
$L$ in contrast to the $q\bar q$ interpretation, can reproduce the $SU(3)_{\rm F}$ nonet structure
and mass ordering in a natural way \cite{Jaffe:1976ig,Hooft:2008we}. The physical
$f_0(980)$ and $\sigma(600)$ states are given in terms of the ideally mixed states
\begin{equation}
|f^{[0]}_0(980)\rangle\equiv\frac{[su][\bar s \bar u]+[sd][\bar s\bar d]}{\sqrt{2}},
\quad
|\sigma^{[0]}(600)\rangle\equiv[ud][\bar u\bar d]
\quad
\end{equation}
as
\begin{equation}\label{tetra}
\left(
\begin{array}{c}
|f_0(980)\rangle\\
|\sigma(600)\rangle
\end{array}
\right)=
\left(
\begin{array}{cc}
\cos\omega\  & \ -\sin\omega \\
\sin\omega\  & \ \cos\omega
\end{array}
\right)
\cdot
\left(
\begin{array}{c}
|f^{[0]}_0(980)\rangle\\
|\sigma^{[0]}(600)\rangle
\end{array}
\right).
\end{equation}
An analysis of the measured scalar masses points to a small deviation from
ideal mixing, with an upper bound of $|\omega|<5^\circ$
\cite{Hooft:2008we,Black:1998wt,Maiani:2004uc}, which we shall neglect in the following
discussion.

In Ref.~\cite{Hooft:2008we}, it was pointed out that a coherent picture of the scalar mesons can be
obtained through mixing between tetraquark and $q\bar q$ states due to instanton effects.
Here the light scalar mesons $\lsim 1$~GeV are predominantly tetraquark states while their heavier
counterparts, with masses $\gsim 1$~GeV, are predominantly $q\bar q$ states.
A fit of this model to data adequately describes the mass spectrum.

\boldmath
\subsection{Further Probes of the $f_0(980)$}
\unboldmath
In lattice QCD, there is an ongoing effort to calculate  the spectrum of the low-lying
scalars and to study observables that will allow us to distinguish between exotic and conventional
states (see, e.g., Ref.~\cite{Alford:2000mm}).

On the phenomenological side,
several processes are under scrutiny to probe the structure of the $f_0(980)$.
Particularly interesting are radiative  $\phi \to f_0 \gamma$ decays, which were proposed
to distinguish between the standard $ q \bar q$ and tetraquark interpretation. One evident
difference is that the radiative transition of the $\phi \sim s \bar s$ to a non-strange
$q \bar q $ state would require the annihilation and creation of an additional quark--antiquark pair
in the $q\bar q$ picture, which is suppressed by the  Okubo--Zweig--Iizuka (OZI) rule. On the
other hand, the transition to a  $ q q \bar q \bar q$ state containing a $s \bar s $ pair
requires only the creation of an additional $q \bar q $ pair, which is not OZI-suppressed.
Also here the data seem to favour a tetraquark picture, although alternative interpretations
involving model-dependent assumptions are possible as well
(see, e.g.,  Refs.~\cite{Achasov:1987ts, Delbourgo:1998gi}).

Further insights come from  two-photon fusion or decays into two photons
thanks to their sensitivity to the electric charge of the constituent quarks of the $f_0(980)$
(see, e.g., Refs.~\cite{Delbourgo:1998gi, barnes}).

The  $f_0(980)$ has also been observed in hadronic decays of $Z^0$ bosons, where its
production properties are found to be similar to those of a $\phi$ meson.
This supports the $q\bar q$ picture although there are currently no predictions for the production
rates of the tetraquark or the even more exotic $K \bar K$ molecule picture available \cite
{Ackerstaff:1998ue}.

Over the last decade, a large amount of data for decays of heavy mesons has become
available, opening the way for new studies to reveal the hadronic structure of the
$f_0(980)$. Decays of $D_s$ mesons have received a lot of attention, and also charmless
hadronic $B$ decays offer a nice laboratory to shed further light on the nature of the scalar
mesons (see, for instance, Refs.~\cite{Deandrea:2000yc, CCY}).

A more comprehensive overview of the hadronic structure of the $f_0(980)$ is beyond the scope
of this paper. In the following discussion of the $B^0_s\to J/\psi f_0$ decay, we shall
consider the quark--antiquark and tetraquark pictures of the $f_0(980)$ as theoretical benchmarks.

\begin{figure}[tbp] 
   \centering
   \includegraphics[width=15cm]{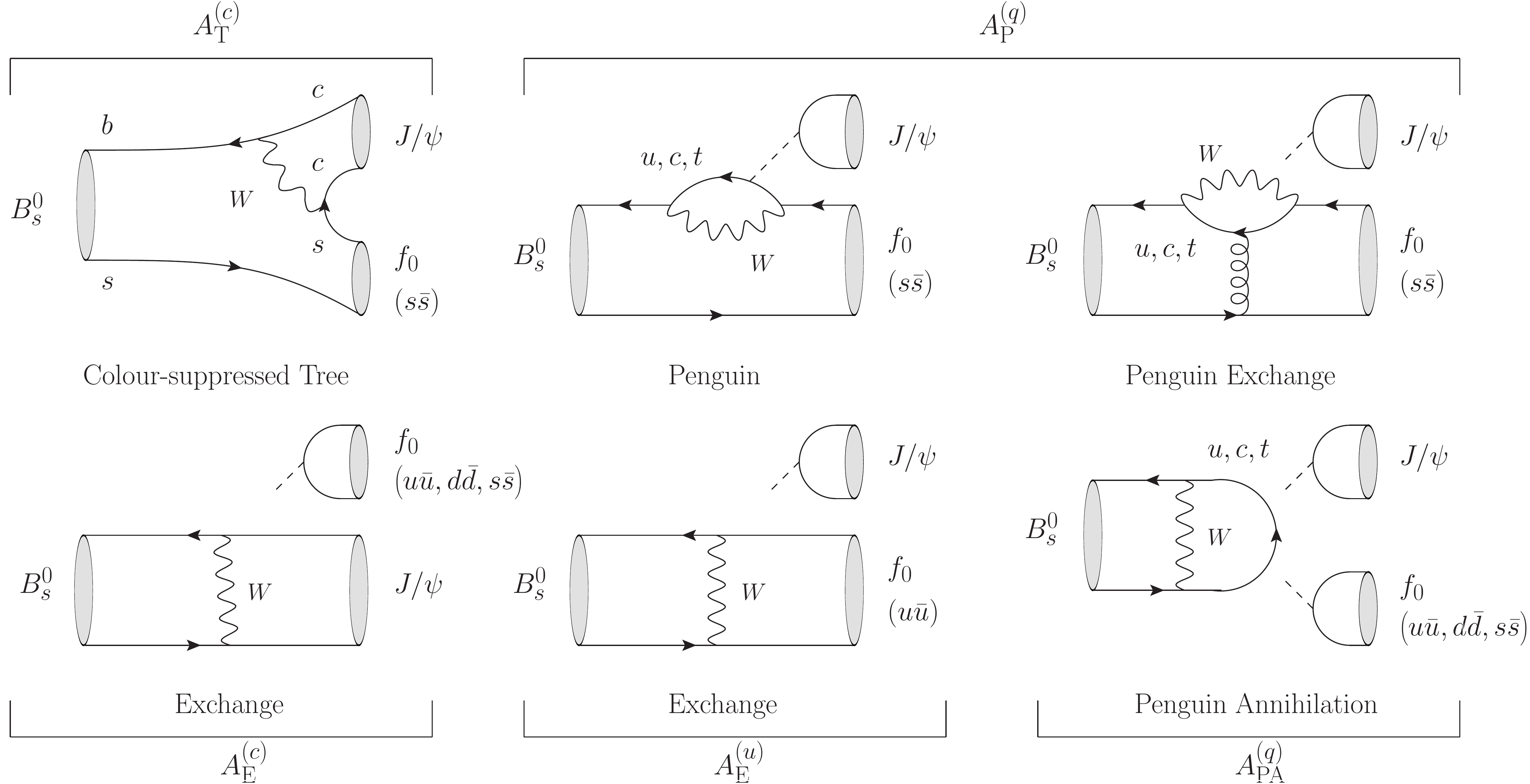}
   \caption{Decay topologies contributing to the $B^0_s\to J/\psi f_0$ channel
   as discussed in the text. The penguin topologies include implicitly
   QCD and EW penguins.}\label{fig:topol}
\end{figure}

\boldmath
\section{Amplitude Structure of $B^0_s\to J/\psi f_0$}\label{sec:ampl}
\unboldmath
\setcounter{equation}{0}
\subsection{Decay Topologies}
In Fig.~\ref{fig:topol}, we show the decay topologies contributing to $B^0_s\to J/\psi f_0$
in the SM. The structure of the corresponding decay amplitude is given as follows:
\begin{align}
A(B^0_s\to J/\psi f_0)=&\ \lambda_c^{(s)}\left[A^{(c)}_{\rm T}+A_{\rm P}^{(c)}
+A_{\rm E}^{(c)} +A_{\rm PA}^{(c)}  \right] + \lambda_u^{(s)} \left[ A_{\rm P}^{(u)}+
A_{\rm E}^{(u)}+A_{\rm PA}^{(u)} \right] \notag\\
&\ +  \lambda_t^{(s)} \left[ A_{\rm P}^{(t)} + A_{\rm PA}^{(t)}\right],
\end{align}
where $\lambda_q^{(s)}\equiv V_{qs} V_{qb}^\ast$ are CKM factors and $A^{(q)}_{\rm topology}$
generically denotes the corresponding CP-conserving strong amplitudes.
Specifically, $A^{(c)}_{\rm T}$ is the colour-suppressed tree contribution, $A_{\rm P}^{(q)}$ are
the penguin and penguin exchange topologies with a $q$-quark running in the
loop, $A_{\rm E}^{(c)}$ and $A_{\rm E}^{(u)}$ describe exchange topologies with
$c\bar c$ and $u\bar u$ pairs created by the $W$ exchange, respectively, while the
$A_{\rm PA}^{(q)}$ denote the penguin annihilation topologies with internal $q$-quarks.

The penguin topologies implicitly include QCD and electroweak (EW) penguins. In analogy to
$B^0_d\to J/\psi K_{\rm S}$ or $B^0_s\to J/\psi \phi$, the QCD penguin topologies
require a colour-singlet exchange and are OZI-suppressed. However, this comment
does not apply to the EW penguin diagrams, which can contribute in colour-allowed form and
are hence expected to have a significant impact on the $B^0_s\to J/\psi f_0$ penguin
sector \cite{RF-habil}.
It is not evident that the OZI suppression is effective for the QCD
penguin topologies and that it cannot be reduced by long-distance effects. Let us also note
that data on non-leptonic $B$-meson decays of the kind $B\to\pi\pi$ and $B\to D\pi$ indicate
that colour suppression is not effective in nature (see, for instance, Refs.~\cite{BFRS,FST-fact}).

Using the unitarity of the CKM matrix to eliminate the $\lambda_t^{(s)}$ factor, we obtain
\begin{equation}\label{ampl-1}
A(B^0_s\to J/\psi f_0)=\left(1-\frac{\lambda^2}{2}\right){\cal A} \left [1+\epsilon b e^{i\vartheta}
e^{i\gamma}  \right],
\end{equation}
where we have introduced the CP-conserving ``hadronic'' parameters
\begin{equation}
{\cal{ A}} \equiv\lambda^2 A \left[A^{(c)}_{\rm T}+A_{\rm P}^{(ct)}+A_{\rm E}^{(c)}
+A_{\rm PA}^{(ct)}\right]\label{Acal}
\end{equation}
and
\begin{equation}
b e^{i\vartheta} \equiv R_b\left[\frac{A_{\rm P}^{(ut)}+
A_{\rm E}^{(u)}+A_{\rm PA}^{(ut)}}{A^{(c)}_{\rm T}+A_{\rm P}^{(ct)}+A_{\rm E}^{(c)}+
A_{\rm PA}^{(ct)}} \right], \label{hadrDefn}
\end{equation}
using the shorthand notation
\begin{equation}
A_{\rm topology}^{(qt)} \equiv A_{\rm topology}^{(q)} - A_{\rm topology}^{(t)},
\end{equation}
with $q\in\{u,c\}$. These CP-conserving amplitudes can be expressed in terms
of hadronic matrix elements of four-quark operators appearing in the relevant
low-energy effective Hamiltonian \cite{RF-habil,BBL}. In the above expressions,
$\lambda\equiv|V_{us}|=0.2252 \pm 0.0009$ is the Wolfenstein
parameter of the Cabibbo--Kobayashi--Maskawa (CKM) matrix \cite{PDG},
\begin{equation} \label{VubVcb}
\epsilon\equiv\frac{\lambda^2}{1-\lambda^2}=0.0534  \pm 0.0005, \quad
A\equiv \frac{|V_{cb}|}{\lambda^2}\sim0.8, \quad
R_b\equiv \left(1-\frac{\lambda^2}{2}\right)\frac{1}{\lambda}\left|\frac{V_{ub}}{V_{cb}}\right|\sim0.5,
\end{equation}
and $\gamma$ is the usual angle of the unitarity triangle (UT) of the CKM matrix. As the parameters
$A$ and $R_b$ will not enter our numerical calculations, we have only indicated their
orders of magnitude. For a detailed discussion of the current experimental information about
$|V_{cb}|$ and $|V_{ub}|$ from semi-leptonic $B$-meson decays,
the reader is referred to Refs.~\cite{PDG,HFAG}.

The form  of the $B^0_s\to J/\psi f_0$ amplitude is similar to that
of $B^0_d\to J/\psi K^0$ \cite{RF-BpsiK,CPS,FFJM} and
$B^0_s\to J/\psi \phi$ \cite{FFM}. In analogy to these channels,
the hadronic parameters $b$ and $\vartheta$ cannot be calculated reliably and suffer
from large theoretical uncertainties. In the case of the
$B^0_s\to J/\psi f_0$ channel the situation is even worse because the details
of the hadronic composition of the $f_0(980)$ affects the value of $be^{i\vartheta}$.
However, the crucial feature is that this parameter enters the decay amplitude with
the tiny $\epsilon$ factor, i.e.\ it is doubly Cabibbo-suppressed.

\begin{figure}[tbp] 
   \centering
   \includegraphics[width=12cm]{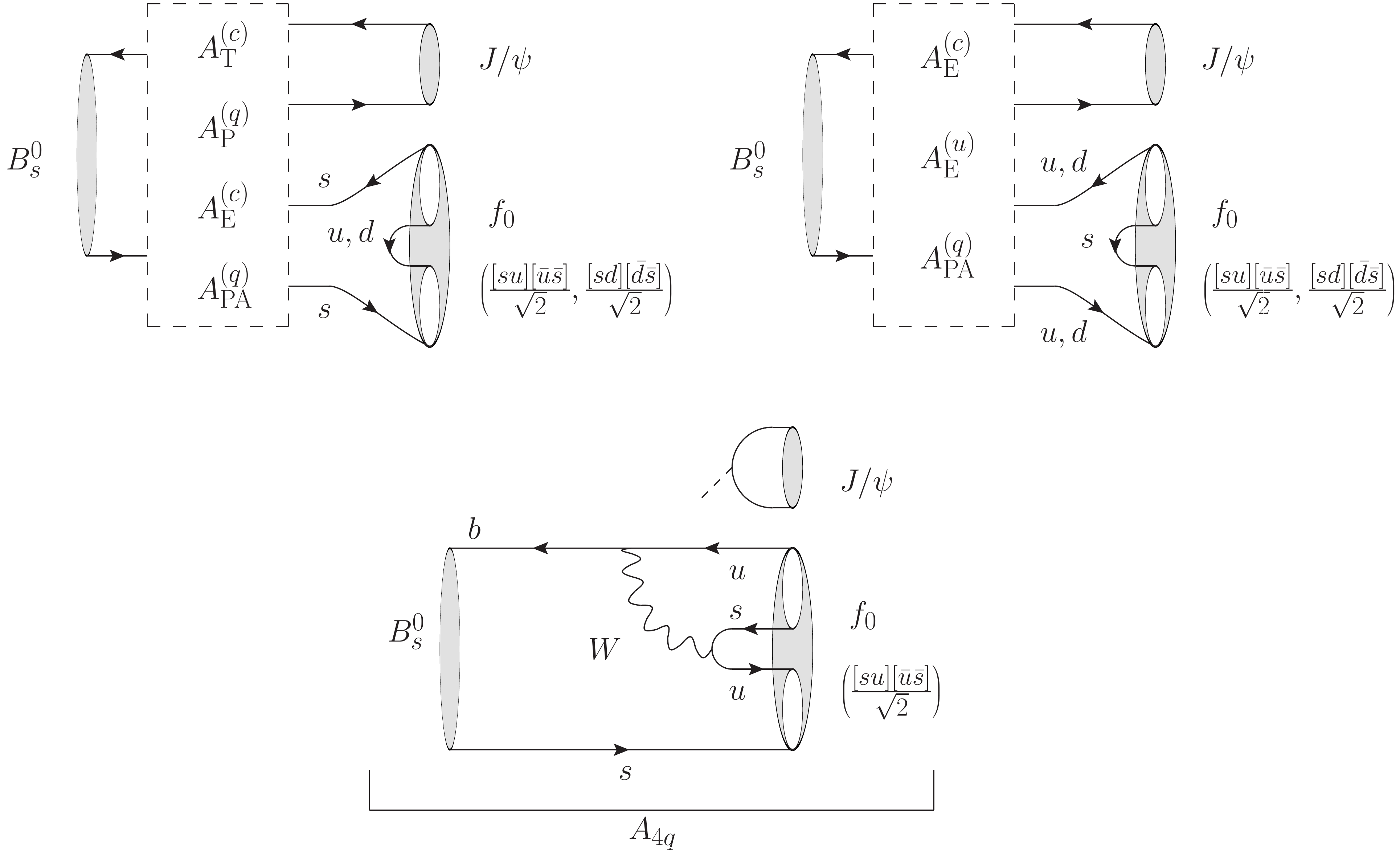}
   \caption{Illustration of how the topologies shown in Fig.~\ref{fig:topol} are extended in the
   case where the $f_0(980)$ is a tetraquark. Also shown is the additional topology
   $A_{4q}$.}\label{fig:f0topol}
\end{figure}

\boldmath
\subsection{Specific Assumptions about the $f_0(980)$}\label{sec:assumptions}
\unboldmath
In order to obtain insights into the parameter $be^{i\vartheta}$, we have to make assumptions
about the hadronic composition of the $f_0(980)$. In the case where the $f_0(980)$ adheres
to the $q\bar{q}$ model as described by \eqref{f0-structure}, we may split the strong amplitudes
into separate terms, projecting out on the different quark flavours. This gives
\begin{align}
A^{(c)}_{\rm T} &= \cos\mixAng \tilde{A}^{(c)}_{{\rm T},s\bar{s}},&
A^{(qt)}_{\rm P} &= \cos\mixAng \tilde{A}^{(qt)}_{{\rm P},s\bar{s}},
\notag\\
A^{(c)}_{\rm E} &= \cos\mixAng \tilde{A}^{(c)}_{{\rm E},s\bar{s}}
	+\frac{1}{\sqrt{2}}\sin\mixAng\left[ \tilde{A}^{(c)}_{{\rm E},u\bar{u}}
	+\tilde{A}^{(c)}_{{\rm E},d\bar{d}} \right], &
A^{(u)}_{\rm E} &=\frac{1}{\sqrt{2}}\sin\mixAng \tilde{A}^{(u)}_{{\rm E},u\bar{u}}, \notag\\
A^{(qt)}_{\rm PA} &= \cos\mixAng \tilde{A}^{(qt)}_{{\rm PA},s\bar{s}}
	+\frac{1}{\sqrt{2}}\sin\mixAng\left[ \tilde{A}^{(qt)}_{{\rm PA},u\bar{u}}
	+\tilde{A}^{(qt)}_{{\rm PA},d\bar{d}} \right], &
\label{BsFlavDecomp}
\end{align}
where $\tilde{A}^{(q')}_{{\rm topology},q\bar{q}}$ denotes a CP-conserving strong
amplitude contributing to the $q\bar{q}$ flavour component of  the $f_0(980)$.
This decomposition is analogous to $SU(3)_{\rm F}$ analyses of non-leptonic
$B$ decays involving $\eta$ or $\eta'$ mesons, where we have to deal
with $\eta$--$\eta'$ mixing \cite{DGR,skands,CGR}.
By assuming $SU(3)_{\rm F}$ flavour symmetry for the strong dynamics producing the
$f_0(980)$, we can, for convenience, drop the $q\bar{q}$ subscripts
without further loss of generality. The hadronic parameter defined in \eqref{hadrDefn} then
takes the following form:
\begin{equation}
\left. b e^{i\vartheta}\right|_{q\bar{q}} = R_b\left[
	\frac{\cos\mixAng\left\{\tilde{A}^{(ut)}_{\rm P} + \tilde{A}^{(ut)}_{\rm PA}\right\} +
	\frac{1}{\sqrt{2}}\sin\mixAng \left\{\tilde{A}^{(u)}_{\rm E} + 2 \tilde{A}^{(ut)}_{\rm PA}\right\} }
	{\cos\mixAng\left\{\tilde{A}^{(c)}_{\rm T} + \tilde{A}^{(ct)}_{\rm P} + \tilde{A}^{(c)}_{\rm E} + \tilde
	{A}^{(ct)}_{\rm PA}\right\}
	+\frac{1}{\sqrt{2}}\sin\mixAng \left\{ 2 \tilde{A}^{(c)}_{\rm E} +2 \tilde{A}^{(ct)}_{\rm PA}\right\}}
	\right].
	\label{btheta}
\end{equation}

If, instead, the $f_0(980)$ is a tetraquark, the $u\bar{u}$, $d\bar{d}$ and $s\bar{s}$ final states
of the topologies in Fig.~\ref{fig:topol} are modified by the creation of an extra quark--antiquark
pair as shown in Fig.~\ref{fig:f0topol}. Moreover, there is
an additional topology $A_{4q}$, which is specific to the tetraquark description of the $f_0(980)$.
Another example of a weak $B$-meson decay with an additional topology in the tetraquark
interpretation of the light scalars that is not present in the $q\bar q$ picture is the
$B^0_d\to \kappa^+K^-$ channel, as was pointed out in Ref.~\cite{CCY}.

In order to simplify the discussion, we assume $\omega=0$ in (\ref{tetra}). The
strong amplitudes can then be written as follows:
\begin{align}
A^{(c)}_{\rm T} &= \frac{1}{\sqrt{2}}\left(\tilde{A}^{(c)}_{{\rm T},su\bar{u}\bar{s}} +
\tilde{A}^{(c)}_{{\rm T},sd\bar{d}\bar{s}}\right)
	\stackrel{\rm isospin}{=} \sqrt{2} \tilde{A}^{(c)}_{{\rm T}},\notag\\
A^{(qt)}_{\rm P} &=  \frac{1}{\sqrt{2}}\left(\tilde{A}^{(qt)}_{{\rm P},su\bar{u}\bar{s}}
+ \tilde{A}^{(qt)}_{{\rm P},sd\bar{d}\bar{s}}\right)
	\stackrel{\rm isospin}{=} \sqrt{2} \tilde{A}^{(qt)}_{{\rm P}},
\notag\\
A^{(c)}_{\rm E} &= \frac{1}{\sqrt{2}}\left(
	  \tilde{A}^{(c)}_{{\rm E},su\bar{u}\bar{s}}
	+ \tilde{A}^{(c)}_{{\rm E},sd\bar{d}\bar{s}}
	+ \tilde{A}^{(c)}_{{\rm E},us\bar{s}\bar{u}}
	+ \tilde{A}^{(c)}_{{\rm E},ds\bar{s}\bar{d}}
	\right)
	\stackrel{SU(3)_{\rm F}}{=} 2\sqrt{2} \tilde{A}^{(c)}_{{\rm E}},\notag\\
A^{(u)}_{\rm E} &=\frac{1}{\sqrt{2}} \tilde{A}^{(u)}_{{\rm E},us\bar{s}\bar{u}}
	= \frac{1}{\sqrt{2}} \tilde{A}^{(u)}_{{\rm E}},\notag\\
A^{(qt)}_{\rm PA} &=  \frac{1}{\sqrt{2}}\left(
	  \tilde{A}^{(c)}_{{\rm PA},su\bar{u}\bar{s}}
	+ \tilde{A}^{(c)}_{{\rm PA},sd\bar{d}\bar{s}}
	+ \tilde{A}^{(c)}_{{\rm PA},us\bar{s}\bar{u}}
	+ \tilde{A}^{(c)}_{{\rm PA},ds\bar{s}\bar{d}}
	\right)
	\stackrel{SU(3)_{\rm F}}{=} 2\sqrt{2} \tilde{A}^{(qt)}_{{\rm PA}},
\label{TetraDecomp}
\end{align}
where $\tilde{A}^{(q')}_{{\rm topology},qq'\bar{q}'\bar{q}}$ denotes a strong amplitude
of which the $f_0(980)$ tetraquark was formed by a $q\bar{q}$ final state (from the corresponding
topology in Fig.~\ref{fig:topol}) combining with a $q'\bar{q}'$ pair.
In the last equalities of the expressions in \eqref{TetraDecomp} we have assumed -- as
indicated -- isospin or $SU(3)_{\rm F}$ symmetry in order to simplify them. The additional
topology $A_{4q}$ in Fig.~\ref{fig:f0topol} can be written correspondingly as
\begin{equation}
A_{4q}=\frac{1}{\sqrt{2}}\tilde A_{4q,us\bar u\bar s}=\frac{1}{\sqrt{2}}\tilde A_{4q},
\end{equation}
and contributes with the CKM factor $\lambda_u^{(s)}$.
We finally arrive at the following
expression for the hadronic parameter defined in
\eqref{hadrDefn}:
\begin{equation}
\left. b e^{i\vartheta}\right|_{\rm 4q} = R_b\left[
	\frac{\tilde{A}^{(ut)}_{\rm P} +
	\frac{1}{2}\tilde{A}^{(u)}_{\rm E} + 2 \tilde{A}^{(ut)}_{\rm PA} +
	\frac{1}{2}\tilde A_{4q}
	 }
	{\tilde{A}^{(c)}_{\rm T} + \tilde{A}^{(ct)}_{\rm P} + 2\tilde{A}^{(c)}_{\rm E} + 2\tilde{A}^{(ct)}_{\rm
	PA}
	}\right].
	\label{bTetra}
\end{equation}
It is interesting to observe that in the absence of the $A_{4q}$ contribution $\left. b e^{i\vartheta}\right|_{q\bar{q}}$ takes the same form as $\left. b e^{i\vartheta}\right|_{\rm 4q}$ for
\begin{equation}\label{phi-4q}
\cos\mixAng=\sqrt{\frac{2}{3}}, \quad
\sin\mixAng=\sqrt{\frac{1}{3}},
\end{equation}
i.e.\ for a mixing angle of $\mixAng=35^\circ$, which corresponds to
\begin{equation}
|f_0(980)\rangle = \frac{1}{\sqrt{6}}\left[ |u\bar u\rangle + |d\bar d\rangle+2|s\bar s\rangle \right],
\end{equation}
with a flavour structure similar to that of the $\eta'$ meson \cite{DiDonato:2011kr,DGR}.
The individual topological amplitudes would, however,
still take different values in the quark--antiquark and tetraquark descriptions of the $f_0(980)$.
Unfortunately, we cannot calculate these amplitudes as they are non-perturbative quantities.

For the discussion of the $B^0_s\to J/\psi f_0$ observables in Sections~\ref{sec:lifetime}
and \ref{sec:CP}, we will consider the following range for the relevant hadronic parameters:
\begin{equation}\label{b-range}
0\leq b \leq 0.5, \quad 0^\circ \leq \vartheta \leq 360^\circ.
\end{equation}
Because of $R_b\sim 0.5$, the value of $b\sim0.5$ would correspond to strong amplitudes
in the numerator and denominator of (\ref{hadrDefn}) of the same order of magnitude.
In view of the still unsettled hadronic structure of the $f_0(980)$ and the complex -- and
essentially unknown -- hadronization dynamics of the $B^0_s\to J/\psi f_0$ channel we
cannot exclude such a scenario.\footnote{Using experimental data on $B^0_d\to J/\psi\pi^0$
and the $SU(3)_{\rm F}$ flavour symmetry, the counterparts of $b$ and $\vartheta$ in
$B^0_d\to J/\psi K^0$, $a$ and $\theta$, are found at the $1\,\sigma$ level as
$a\in[0.15,0.67]$ and $\theta\in[174,212]^\circ$ \cite{FFJM}.} We shall return to the
hierarchy of the different decay topologies in Section~\ref{sec:hier}.

The expressions for $be ^{i\vartheta}$ in the $q\bar{q}$ and tetraquark pictures will be useful
when discussing the $B_d^0\to J/\psi f_0(980)$ channel in Section~\ref{sec:Bd}.

\subsection{Estimate of the Branching Ratio in Factorization}\label{sec:BRest}
It is instructive to estimate the branching ratio of $B^0_s\to J/\psi f_0$ from the measured
$B^0_d\to J/\psi K^0$ branching ratio. To this end, we assume that the $f_0(980)$ is
a quark--antiquark state satisfying (\ref{f0-structure}). In the factorization
approximation, the hadronic matrix element takes the following form \cite{col}:
\begin{eqnarray}
\langle f_0(p')|\bar s \gamma^\mu\gamma_5 b|B^0_s(p)\rangle &=& \cos\mixAng F_{1,{s\bar{s}}}^
{B^0_sf_0}(q^2)
\left [(p+p')^\mu-\left(\frac{M_{B^0_s}^2-M_{f_0}^2}{q^2}\right)q^\mu \right]\nonumber\\
&&+\cos\mixAng F_{0,{s\bar{s}}}^{B^0_s f_0}(q^2)\left(\frac{M_{B^0_s}^2-M_{f_0}^2}{q^2}\right)q^
\mu,\label{FF-A}
\end{eqnarray}
where $q\equiv p-p'$ is the transfered momentum and $M$ denotes a particles mass.
Here the $F_{k,{s\bar{s}}}^{B^0_sf_0}$, with $k\in\{1,2\}$, are form factors that describe the
transition of the $B_s^0$ meson to the $s\bar{s}$ component of the $f_0(980)$.
Since the $f_0(980)$ is a scalar particle,
Lorentz invariance implies that only the axial-vector part of the $V-A$ current contributes
to the matrix element with the pseudo-scalar $ B^0_s$ meson. In the case of the
$B^0_d\to J/\psi K^0$ transition, as also the $K^0$ is a pseudoscalar meson, only
the matrix element of the vector current $\bar s \gamma^\mu b$ is non-vanishing,
with a parametrization in terms of form factors that is completely analogous to (\ref{FF-A}).

If we use the factorization approximation and take only the colour-suppressed
tree-diagram-like topology $A^{(c)}_{\rm T}$ into account, we obtain
\begin{equation}
\left.\frac{\mbox{BR}(B_s^0\to J/\psi f_0)}{\mbox{BR}(B_d^0\to J/\psi K^0)}\right|_{\rm fact.}=
\frac{\tau_{B^0_s}}{\tau_{B^0_d}}\left(\frac{M_{B^0_s}\Phi_s}{M_{B^0_d}\Phi_d}\right)^3
\left[\frac{F_{1,{s\bar{s}}}^{B^0_sf_0}(M_{J/\psi}^2)}{F_1^{B^0_dK^0}(M_{J/\psi}^2)}
\right]^2 \cos^2\mixAng ,
\end{equation}
where
\begin{equation}
\Phi_s\equiv \Phi(M_{J/\psi}/M_{B^0_s},M_{f_0}/M_{B^0_s}), \quad
\Phi_d\equiv \Phi(M_{J/\psi}/M_{B^0_d},M_{K^0}/M_{B^0_d})
\end{equation}
with
\begin{equation}
\Phi(x,y)\equiv \sqrt{\left[1-\left(x+y\right)^2\right]\left[1-\left(x-y\right)^2\right]}
\end{equation}
are phase-space factors.

To calculate the $B^0_d\to K^0$ form factor we use the leading-order light-cone QCD sum-rule
analysis of Ref.~\cite{Ball:2004ye} and extrapolate to the scale of interest, $M_{J/\psi}^2$, by
using the analytic evolution equations in $q^2$ provided. The resulting value  is very similar
to the one that can be inferred from the plots of Ref.~\cite{DuMe};  we obtain
\begin{equation}
F_1^{B^0_dK^0}(M_{J/\psi}^2) = 0.615\pm{0.076}.
\label{FF2}
\end{equation}

The $B^0_s\to f_0(980)$ form factor of the axial-vector current is more problematic due to the
uncertain mixing angle $\mixAng$.
The authors of Ref.~\cite{col} perform a leading order light-cone QCD sum-rule
calculation with the assumption that the $f_0(980)$ is entirely an $s\bar{s}$ state, i.e.\
that $\mixAng = 0^\circ$. Using the evolution equation in $q^2$ that they provide, we obtain
\begin{equation}\label{FF1a}
\left[\cos\mixAng F_{1,s\bar{s}}^{B^0_sf_0}(M_{J/\psi}^2)\right]_{\mixAng =
0^\circ} = 0.32^{+0.06}_{-0.05}.
\end{equation}

On the other hand, the authors of Ref.~\cite{ElBennich:2008xy} determine the mixing angle
$\mixAng$ by performing a fit of the $D_{(s)}\to f_0(980)\{\pi, K\}$ branching ratios with the
help of two approaches: covariant light-front dynamics and dispersion relations. The latter
method gives a fitted mixing angle of $\mixAng = 41.6^\circ$ and is better behaved for the large
momentum transfer $M^2_{J/\psi}$; we read off from their plot:
\begin{equation}\label{FF1b}
\left[\cos\mixAng F_{1,s\bar{s}}^{B^0_sf_0}(M_{J/\psi}^2)\right]_{\mixAng = 41.6^\circ} \simeq 0.5.
\end{equation}
Due to the wide variance in results for the different methods and mixing angles, we will not include
error estimates for the corresponding branching ratio calculation.

By combining the form factors in \eqref{FF2}, \eqref{FF1a} and \eqref{FF1b} with the measured
value $\mbox{BR}(B_d^0\to J/\psi K^0)=(8.71\pm0.32)\times10^{-4}$, as well as the lifetimes and
mass values listed in Ref.~\cite{PDG}, we find
\begin{align}
\left.\mbox{BR}(B_s^0\to J/\psi f_0)\right|_{\mixAng = 0^\circ}&\simeq 1.9\times10^{-4}
\label{est-1}
\intertext{and}
\left.\mbox{BR}(B_s^0\to J/\psi f_0)\right|_{\mixAng = 41.6^\circ}&\simeq 4.8\times10^{-4}.
\label{eqn:BrBf0-FF}
\end{align}
Let us emphasize that these estimates assume that the $f_0(980)$ is described in the $q\bar q$
picture by (\ref{f0-structure}), include only tree topologies, and take only factorizable
$SU(3)$-breaking effects through the form-factor calculations listed above into account.

\subsection{Estimate of the Branching Ratio from Experiment}
We proceed to compare the results obtained in the previous subsection with
the data listed in Table~\ref{tab:BRs}. Combining errors in quadrature and taking a weighted
average gives\footnote{Our simple weighted average does not account for possible correlations
between uncertainties. We hope that more sophisticated averages will be available from the
Particle Data Group soon \cite{PDG}.}
\begin{equation}
\left.{\rm BR}(B_s^0 \to J/\psi f_0; f_0 \to \pi^+ \pi^-)\right|_{\rm avg} = \left(1.36^{+0.28}_{-0.24}\right)
\times 10^{-4} .
\label{eqn:Bf0PiPi}
\end{equation}
The missing ingredient is ${\rm BR}(f_0 \to \pi^+ \pi^-)$, which has not been adequately measured.
However, measurements do exist for the ratios
\begin{equation}
R \equiv \frac{\Gamma(f_0\to \pi\pi)}{\Gamma(f_0\to\pi\pi) + \Gamma(f_0\to KK)}\quad{\rm and}\quad
R'\equiv\frac{\Gamma(f_0\to K^+ K^-)}{\Gamma(f_0\to\pi^+\pi^-)}.
\end{equation}
Under the assumption that all other decay channels (such as $\gamma\gamma$) are neglegible
and that the $\pi\pi$ and $KK$ channels adhere to isospin symmetry, we expect
\begin{equation} \label{RR}
{\rm BR}(f_0 \to \pi^+ \pi^-) = \frac{2R}{3} = \frac{2}{4R'+3}
\end{equation}
as well as
\begin{equation}  \label{RR1}
{\rm BR}(f_0 \to K^+ K^-) = \frac{1}{2}\left(1-R\right) = \frac{2 R'}{4R'+3},
\end{equation}
which we include for completeness.
We note, however, that the isospin assumption $\Gamma(f_0\to K^+ K^-)=\Gamma(f_0\to K^0
\bar K^0)$ on which \eqref{RR1} and the expressions involving $R'$ depend, could be spoiled
by phase-space effects.
Specifically, because the decay thresholds of both the $f_0\to K^+ K^-$ and $f_0\to
K^0 \bar K^0$ channels are beyond the $f_0$ mass peak, the slope of its wide resonance could
significantly break the isospin assumption.
As the $\pi\pi$ final states have a much lower threshold and thus access to a large phase space,
we expect the equality in \eqref{RR} involving $R$ to be stable under the above considerations.

Using the measurement $R = 0.75^{+0.11}_{-0.13}$, which was reported by BES2 in 2005
\cite{Ablikim:2005kp}, the authors of Ref.~\cite{ElBennich:2008xy}  have used the above relations
to extract
\begin{equation}\label{eqn:f0PiPi}
{\rm BR}(f_0\to\pi^+\pi^-)=0.50^{+0.07}_{-0.09}
\end{equation}
and ${\rm BR}(f_0\to K^+K^-)=0.125^{+0.055}_{-0.065}$.
Using the 2006 BaBar result
$R'= 0.69\pm0.32$~\cite{Aubert:2006nu}, we find ${\rm BR}(f_0\to\pi^+\pi^-)=0.35\pm 0.08$
and ${\rm BR}(f_0\to K^+K^-)=0.24 \pm 0.06$.
Because of the near 1\,$\sigma$ discrepancy of these results and the preceding discussion
concerning the possible $f_0 \to K\bar{K}$ isospin-breaking effects, we do not use the latter result.

By na\"{\i}vely assuming a narrow width for the $f_0$, we can combine the average in \eqref
{eqn:Bf0PiPi} with \eqref{eqn:f0PiPi} to finally obtain
\begin{equation}
\left.{\rm BR}(B_s^0 \to J/\psi f_0)\right|_{\rm exp} = \left(2.7^{+0.7}_{-0.6}\right)\times 10^{-4},
\end{equation}
which agrees within the errors with the estimates in (\ref{est-1}) and (\ref{eqn:BrBf0-FF}).
Thus the colour-suppressed tree topologies by themselves account for the correct order
of magnitude of the measured $B^0_s\to J/\psi f_0$ branching ratio in the
quark--antiquark picture. However, in view of the large errors, we cannot draw further
conclusions about the hadronic structure of the $f_0(980)$ from this exercise.

\boldmath
\section{Effective Lifetime of $B^0_s\to J/\psi f_0$}\label{sec:lifetime}
\unboldmath
\setcounter{equation}{0}
\boldmath
\subsection{Untagged Decay Rate}
\unboldmath
From an experimental point of view, ``untagged" studies of $B_s$-meson decays, where
we do not distinguish between initially, i.e.\ at time $t=0$, present $B^0_s$ or $\bar B^0_s$
states, are more accessible. Since both $B^0_s$ and $\bar B^0_s$ can decay into the
$J/\psi f_0(980)$ final state, we obtain \cite{DF}
\begin{align}
	\langle \Gamma(B_s(t)\to J/\psi f_0)\rangle
	\equiv&\ \Gamma(B^0_s(t)\to J/\psi f_0)+ \Gamma(\bar B^0_s(t)\to J/\psi f_0)
	\notag\\
	=&\ R_{\rm H}(B_s\to J/\psi f_0)e^{-\Gamma_{\rm H}^{(s)} t} +  R_{\rm L}(B_s\to J/\psi f_0)
	e^{-\Gamma_{\rm L}^{(s)} t}\label{untagged}\\
	\propto &\ e^{-\Gamma_st}\left[ \cosh\left(\frac{\Delta\Gamma_s t}{2}\right)+
	{\cal A}_{\rm \Delta\Gamma}(B_s\to J/\psi f_0)\,\sinh\left(\frac{\Delta\Gamma_s t}
	{2}\right)\right],\notag
\end{align}
where
\begin{equation}
\Delta\Gamma_s\equiv \Gamma_{\rm L}^{(s)}-\Gamma_{\rm H}^{(s)}, \quad
\Gamma_s\equiv \frac{\Gamma_{\rm H}^{(s)}+\Gamma_{\rm L}^{(s)}}{2}=\tau_{B_s}^{-1},
\end{equation}
with $\tau_{B_s}$ denoting the $B^0_s$ lifetime, and
\begin{equation}
\Adel  \equiv {\cal A}_{\Delta\Gamma}(B_s\to J/\psi f_0)\equiv
\frac{R_{\rm H}(B_s\to J/\psi f_0)-R_{\rm L}(B_s\to J/\psi f_0)}{R_{\rm H}(B_s\to J/\psi f_0)+
R_{\rm L}(B_s\to J/\psi f_0)}.
\end{equation}
Using the standard $B^0_s$--$\bar B^0_s$ mixing formalism \cite{RF-habil},
${\cal A}_{\Delta\Gamma}$ is given as follows:
\begin{equation}
{\cal A}_{\Delta\Gamma}=\frac{2\,\mbox{Re}\,\xi_{J/\psi f_0}^{(s)}}{1+
\bigl|\xi_{J/\psi f_0}^{(s)}\bigr|^2},
\end{equation}
where
\begin{equation}\label{xi-expr}
\xi_{J/\psi f_0}^{(s)}=-\eta_{J/\psi f_0}e^{-i\phi_s}\left[ \frac{1+\epsilon b e^{i\vartheta}
e^{-i\gamma} }{1+\epsilon b e^{i\vartheta}
e^{+i\gamma} }\right].
\end{equation}
Here $\eta_{J/\psi f_0}=-1$ is the CP eigenvalue of the $J/\psi f_0(980)$ final state, while
\begin{equation}\label{phis}
\phi_s\equiv\phi_s^{\rm SM}+\phi_s^{\rm NP}
\end{equation}
denotes the CP-violating $B^0_s$--$\bar B^0_s$ mixing phase.
The SM piece takes the following tiny value \cite{Charles:2011va}:
\begin{equation}\label{eqn:phiSM}
\phi_s^{\rm SM}\equiv-2\beta_s=-2\lambda^2\eta=-(2.08\pm0.09)^\circ,
\end{equation}
while $\phi_s^{\rm NP}$ describes the impact of CP-violating NP contributions to 
$B^0_s$--$\bar B^0_s$ mixing. 

In writing \eqref{xi-expr}, we have adopted the parameterization for the SM decay 
amplitude in \eqref{ampl-1}, thereby making the plausible assumption that 
there is no significant NP contribution at the decay amplitude level. Experimental 
evidence in support of this assumption is given by the absence of large direct CP violation in the 
$B^0_d \to J/\psi K^0$ and $B^+\to J/\psi K^+$ channels \cite{HFAG,PDG}. Putting 
exchange and penguin annihilation topologies aside, these channels emerge from the 
same quark-level transitions as $B^0_s \to J/\psi f_0$.

In (\ref{xi-expr}), we have used (\ref{ampl-1}) and have taken
the CP-odd eigenvalue of the
$J/\psi f_0(980)$ final state into account. Following Ref.~\cite{FFM}, we introduce a
quantity $\Delta\phi$ through
\begin{equation}\label{sDelphi}
\sin\Delta\phi = \frac{2 \epsilon b\cos\vartheta \sin\gamma+\epsilon^2b^2
\sin2\gamma}{N\sqrt{1-C^2}}
\end{equation}
\begin{equation}\label{cDelphi}
\cos\Delta\phi=\frac{1+ 2 \epsilon b\cos\vartheta \cos\gamma+\epsilon^2 b^2
\cos2\gamma}{N\sqrt{1-C^2}}.
\end{equation}
Here
\begin{equation}\label{ACPdir}
C\equiv C(B_s\to J/\psi f_0)=\frac{-2 \epsilon b\sin\vartheta \sin\gamma}{N}
\end{equation}
with
\begin{equation}
N\equiv1+2\epsilon b\cos\vartheta\cos\gamma+\epsilon^2 b^2
\end{equation}
describes the direct CP violation in $B^0_s\to J/\psi f_0$ as we will see in Section~\ref{sec:CP}.
Consequently, we have
\begin{equation}\label{tDelphi}
\tan\Delta\phi=\frac{2 \epsilon b\cos\vartheta \sin\gamma+\epsilon^2b^2
\sin2\gamma}{1+ 2 \epsilon b\cos\vartheta\cos\gamma+\epsilon^2b^2\cos2\gamma}.
\end{equation}
Using only this expression would result in a two-fold ambiguity for $\Delta\phi$. However,
this can be lifted thanks to the information about $\sin\Delta\phi$ and $\cos\Delta\phi$ in
(\ref{sDelphi}) and (\ref{cDelphi}), respectively.
The hadronic phase $\Delta\phi$ allows us to write ${\cal A}_{\Delta\Gamma}$ in the following
compact form:
\begin{equation}\label{ADG}
{\cal A}_{\Delta\Gamma} =\sqrt{1-C^2}
\cos(\phi_s+\Delta\phi).
\end{equation}

\boldmath
\subsection{Calculation of the Effective Lifetime}
\unboldmath
A particularly nice and simple observable that is offered by the $B^0_s\to J/\psi f_0$ decay
is its effective lifetime, which is defined through the following expression \cite{FK}:
\begin{equation}\label{lifetime-def}
 \tau_{J/\psi f_0} \equiv \frac{\int^\infty_0 t\ \langle \Gamma(B_s(t)\to J/\psi f_0)\rangle\ dt}
  {\int^\infty_0 \langle \Gamma(B_s(t)\to J/\psi f_0)\rangle\ dt}.
\end{equation}
This quantity is also the resulting lifetime if the untagged rate with the two exponentials
in (\ref{untagged}) is fitted to a single exponential \cite{DFN,HM}. Using the two-exponential
form in (\ref{untagged}) with $ R_{\rm H,L}\equiv  R_{\rm H,L}(B_s\to J/\psi f_0)$ yields
\begin{equation}
  \tau_{J/\psi f_0} =\frac{ R_{\rm L}/\Gamma_{\rm L}^{(s)2}+ R_{\rm H}/\Gamma_{\rm H}^{(s)2}}
  {R_{\rm L}/\Gamma_{\rm L}^{(s)}+ R_{\rm H}/\Gamma_{\rm H}^{(s)}},
\end{equation}
which can be written in terms of
\begin{equation}
y_s\equiv\frac{\Delta\Gamma_s}{2\Gamma_s},
\end{equation}
the observable $\Adel$ and the $B^0_s$ lifetime as
\begin{equation}	\label{eqn:lifetime}
	\frac{\tau_{J/\psi f_0}}{\tau_{B_s}}
	=\frac{1}{1-y_s^2} \left[\frac{1+2\Adel y_s+y_s^2}{1+ \Adel y_s}\right].
\end{equation}
The first measurement of the effective $B^0_s\to J/\psi f_0$ lifetime has recently been
performed by the CDF collaboration \cite{CDF-f0}, with the following result:
\begin{equation}\label{CDF-res}
\tau_{J/\psi f_0}=\left[1.70^{+0.12}_{-0.11}\,({\rm stat})\pm0.03\,({\rm syst})\right]\mbox{ps}.
\end{equation}

The observable $\Adel$ satisfies the inequality:
\begin{equation}\label{ADG-bound}
-1\leq  \Adel \leq +1.
\end{equation}
Consequently, the general expression in (\ref{eqn:lifetime}) implies the constraints:
\begin{equation}
\frac{1}{1+|y_s|}  \leq \frac{\tau_{J/\psi f_0}}{\tau_{B_s}}  \leq \frac{1}{1-|y_s|}.
\end{equation}
These inequalities rely only on (\ref{ADG-bound}) which can neither be spoiled by
hadronic SM contributions to the decay amplitude nor by NP contributions. Moreover,
if we assume that NP can only affect $\Delta\Gamma_s$ through $B^0_s$--$\bar B^0_s$
mixing, which is a very plausible assumption, we have \cite{Grossman}
\begin{equation}\label{ys}
y_s=\frac{\Delta\Gamma_s^{\rm SM}\cos\tilde\phi_s}{2\Gamma_s}=
y_s^{\rm SM}\cos\tilde\phi_s,
\end{equation}
where
\begin{equation}
\tilde\phi_s\equiv\tilde\phi_s^{\rm SM}+\phi_s^{\rm NP}.
\end{equation}
In the latter expression, $\phi_s^{\rm NP}$ is the NP $B^0_s$--$\bar B^0_s$ mixing phase,
which also enters the phase $\phi_s$ defined in (\ref{phis}) on which $\Adel$ depends, whereas the SM piece
takes the following value \cite{NL}:
\begin{equation}\label{tildephis}
\tilde\phi_s^{\rm SM}=(0.22\pm0.06)^\circ.
\end{equation}
Using (\ref{ys}), we obtain
\begin{equation}\label{bound}
\frac{1}{1+y_s^{\rm SM}}\leq\frac{1}{1+y_s^{\rm SM}|\cos\tilde\phi_s|}
\leq \frac{\tau_{J/\psi f_0}}{\tau_{B_s}}  \leq \frac{1}{1-y_s^{\rm SM}|\cos\tilde\phi_s|}
\leq\frac{1}{1-y_s^{\rm SM}}.
\end{equation}
These inequalities hold for the effective lifetime (as defined in (\ref{lifetime-def}))
of any $B_s\to f$ decay where both $B^0_s$ and $\bar B^0_s$ mesons can decay
into the same final state $f$.

\begin{figure}[t]
   \centering
    \includegraphics[width=8.8truecm]{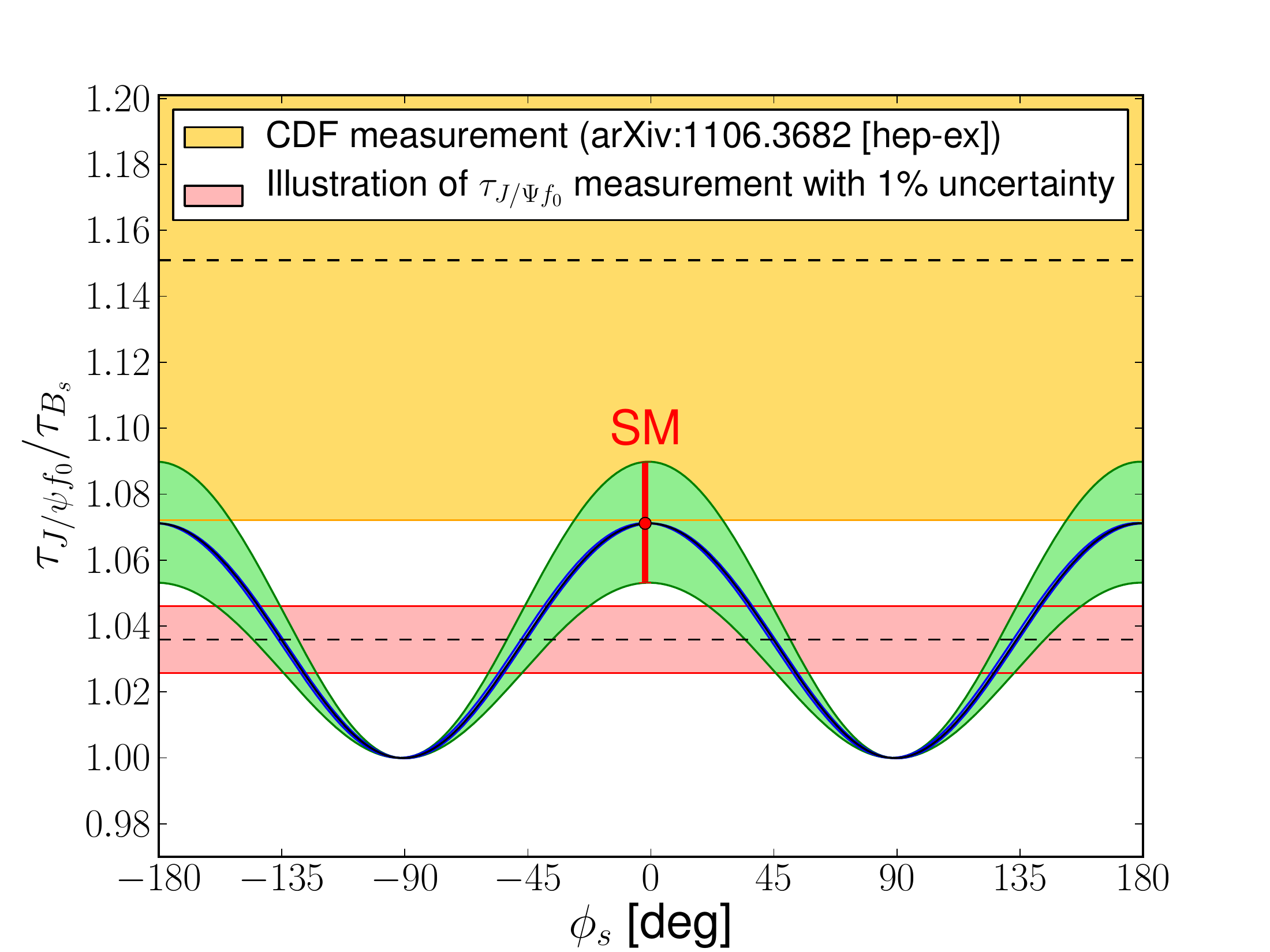}
   \caption{The effective $B^0_s\to J/\psi f_0$ lifetime as a function of the
   $B^0_s$--$\bar B^0_s$ mixing phase $\phi_s$. Assuming $\gamma=(68\pm7)^\circ$ with
   $0\leq b \leq 0.5$ and $0^\circ\leq \vartheta\leq360^\circ$ results in the narrow band in the
   centre of the curve. The major source of the theoretical error comes from the value of
   $\Delta\Gamma_s^{\rm SM}/\Gamma_s=0.133\pm0.032$, as illustrated by the
   wide band of the curve.}\label{fig:lifetime}
\end{figure}

\subsection{Numerical Analysis}
In order to perform a numerical analysis of the effective $B^0_s\to J/\psi f_0$ lifetime, we use
the most recent update for the theoretical analysis of the width difference of the $B_s$-meson
system \cite{NL}:
\begin{equation}\label{DG-SM}
\frac{\Delta\Gamma_s^{\rm SM}}{\Gamma_s}
=0.133\pm0.032.
\end{equation}

In Fig.~\ref{fig:lifetime}, we show the dependence of $\tau_{J/\psi f_0}/\tau_{B_s}$ on the
$B^0_s$--$\bar B^0_s$ mixing phase $\phi_s$. In order to illustrate the impact of the
hadronic corrections, we use
\begin{equation}\label{gam-range}
\gamma=(68\pm7)^\circ,
\end{equation}
which is in agreement with the determination of this angle in Ref.~\cite{FK} and the fits of the
UT \cite{CKMfitter,UTfit}.
For the hadronic parameters we use the ranges in (\ref{b-range}).
Thanks to the suppression from the $\epsilon$ factor in (\ref{ampl-1}),
even this generous
range for $b$ has a very small impact on the effective lifetime. This remarkably robust
behaviour is analogous to the effective lifetimes of other channels, such as
$B^0_s\to K^+K^-$ \cite{FK} and $B^0_s\to J/\psi K_{\rm S}$ \cite{DFK}.

In Fig.~\ref{fig:lifetime}, we also show the CDF measurement given in (\ref{CDF-res}) as the top
horizontal band (the central value is indicated by the dashed line), and observe that it is about
$1\sigma$ above the upper bound for the lifetime, which is numerically governed by the
SM value for $\Delta\Gamma_s/\Gamma_s$ in (\ref{DG-SM}). Our SM prediction of the lifetime
is:
\begin{equation}
\left.\tau_{J/\psi f_0}\right|_{\rm SM} = (1.582\pm 0.036)\,{\rm ps},
\end{equation}
where we have also used (\ref{eqn:phiSM}), (\ref{tildephis}) and
$\tau_{B_s}=(1.477^{+0.021}_{-0.022})\,{\rm ps}$~\cite{HFAG}.
As can be seen in Fig.~\ref{fig:lifetime}, the measurement of $\tau_{J/\psi f_0}$ offers an
interesting probe for CP-violating NP contributions to $B^0_s$--$\bar B^0_s$ mixing. The
lower horizontal band in Fig.~\ref{fig:lifetime} illustrates the impact of a future measurement
of $\tau_{J/\psi f_0}/\tau_{B_s}$ at the $1\%$ level, assuming a value of $\phi_s=-45^\circ$.
It is clearly an important goal to push the measurement of the effective
$B^0_s\to J/\psi f_0$ lifetime to the $1\%$ level.

Since a couple of years, measurements of CP violation in $B^0_s\to J/\psi \phi$ at the
Tevatron indicate possible NP effects in $B^0_s$--$\bar B^0_s$ mixing
\cite{CDF-phis,D0-phis,TEV-LP}. The current status can be summarized as
follows \cite{TEV-LP}: CDF finds the (68\% C.L.) range
\begin{equation}
\phi_s\in[-177.6^\circ,-123.8^\circ] \lor [-59.6^\circ,-2.3^\circ],
\end{equation}
while the \D0 Collaboration reports
\begin{equation}
\phi_s=-\left(31.5^{+20.6}_{-21.8}\right)^\circ.
\end{equation}
These results are complemented by the measurement of the anomalous like-sign dimuon
charge asymmetry at \D0, which was found to differ by $3.9\,\sigma$ from the SM prediction
\cite{di-muon}. The LHCb Collaboration has now also joined the arena, presenting
the currently most precise measurement of $\phi_s$ from the $B^0_s\to J/\psi \phi$
channel \cite{LHCb-f0-CP}:
\begin{equation}\label{LHCb-phis}
\phi_s=+(7.4\pm10.3\pm4.0)^\circ.
\end{equation}
The central value has a sign different from the Tevatron picture and the SM value of $\phi_s$.
Despite tremendous progress, the errors are still sizable and it will be very interesting to
monitor the future measurements. Also the CP violation in $B^0_s\to J/\psi f_0$ provides
information about $\phi_s$, which is our next topic.

\begin{figure}[t]
   \centering
   \begin{tabular}{cc}
     	  \includegraphics[width=7.9truecm]{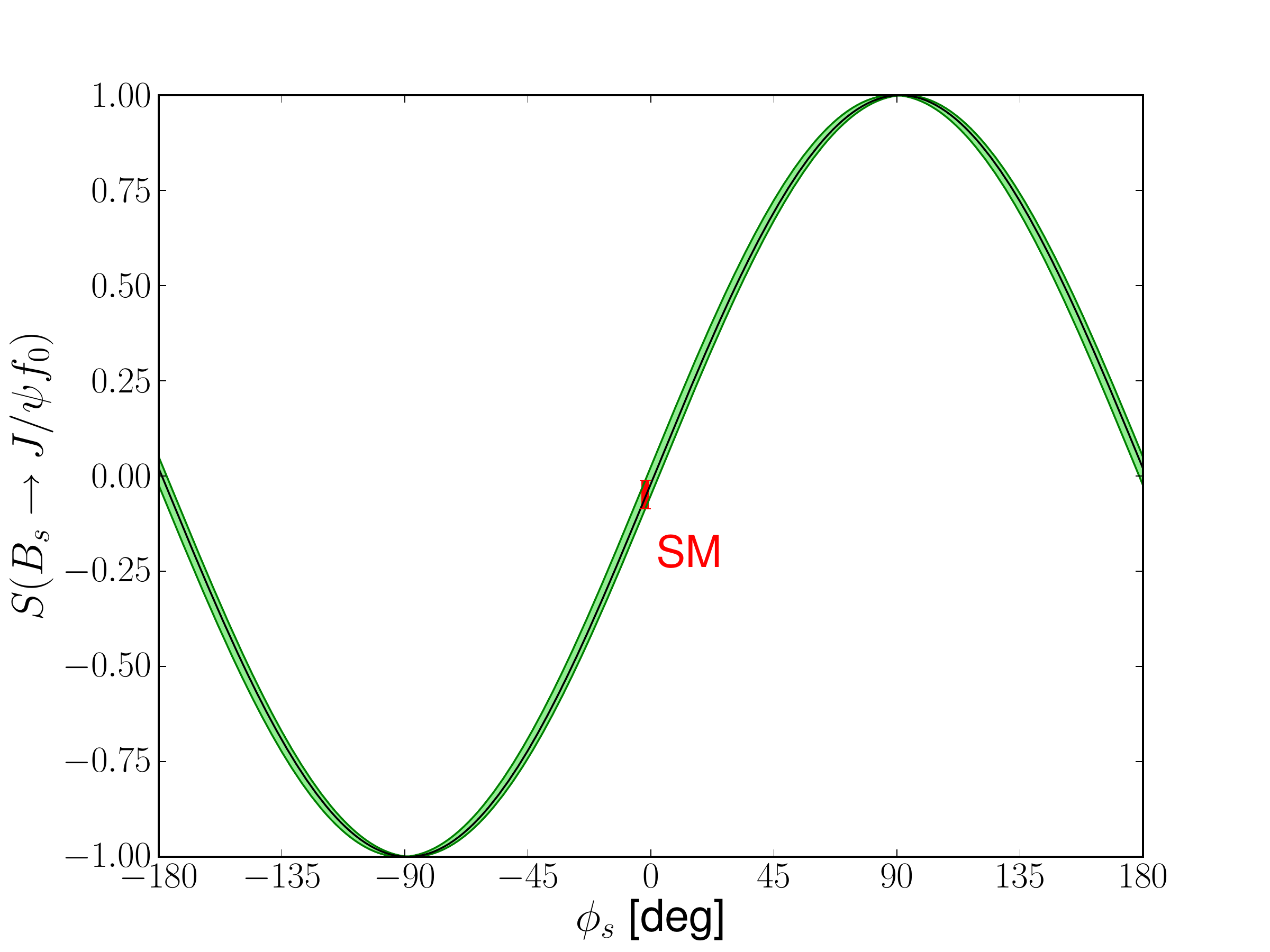} &
        \includegraphics[width=7.9truecm]{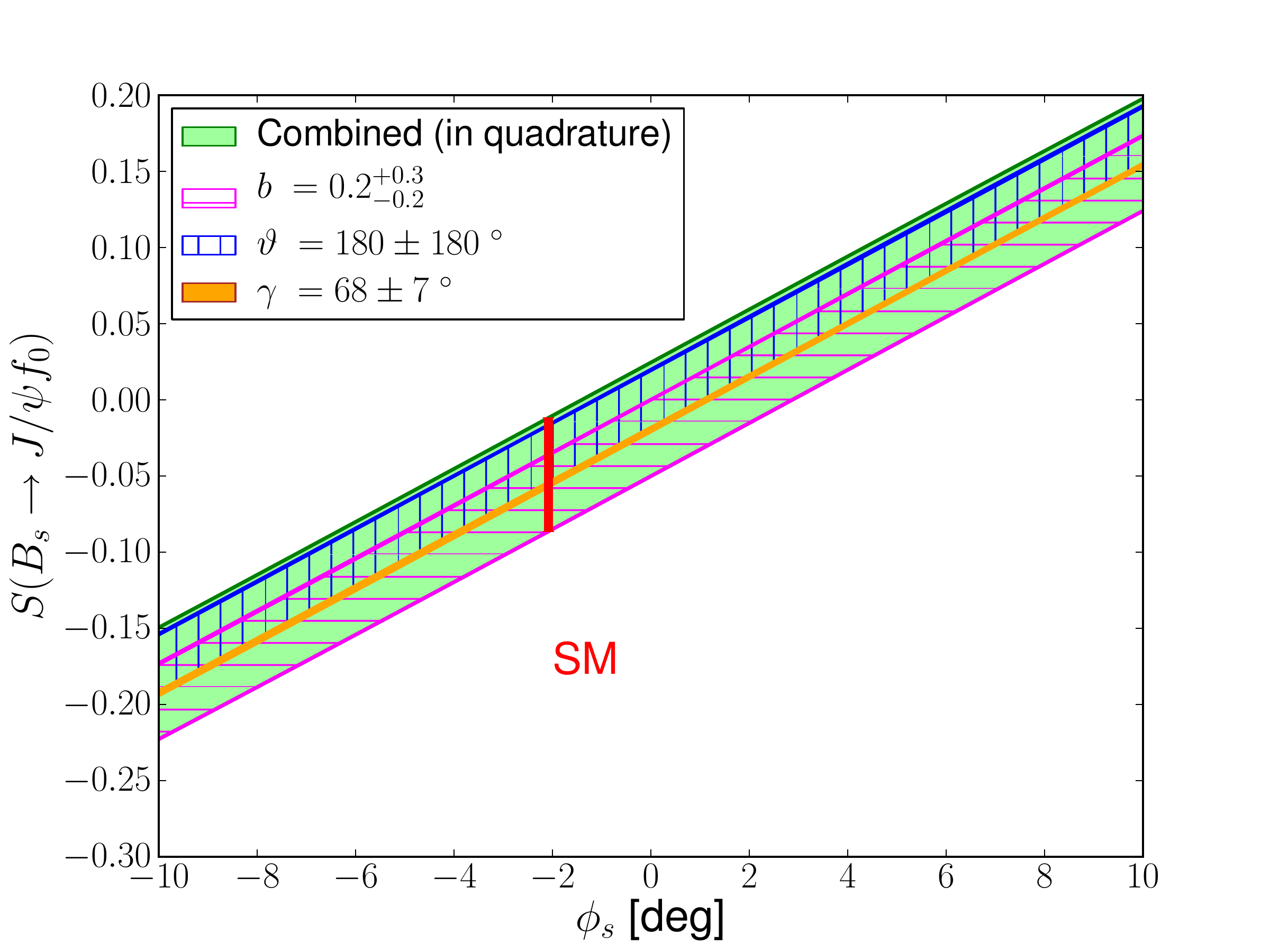}
   \end{tabular}
   \caption{ {\it Left panel}: the mixing-induced CP asymmetry of $B^0_s\to J/\psi f_0$
   as a function of the $B^0_s$--$\bar B^0_s$ mixing phase $\phi_s$;
   assuming $\gamma=(68\pm7)^\circ$, $0\leq b \leq 0.5$ and
   $0^\circ\leq \vartheta\leq360^\circ$ gives the error band.
   {\it Right panel}: individual errors associated with the input quantities, zoomed in on
   the region $\phi_s\in[-10^\circ,10^\circ]$ close to the SM case.
   }\label{fig:CP}
\end{figure}

\boldmath
\section{CP Asymmetries of $B^0_s\to J/\psi f_0$}\label{sec:CP}
\unboldmath
\setcounter{equation}{0}
A tagged analysis, from which we can distinguish between initially present
$B^0_s$ or $\bar B^0_s$ mesons, allows us to measure the  time-dependent,
CP-violating rate asymmetry
\begin{equation}\label{t-dep-asym}
\frac{\Gamma(B_s(t)\to J/\psi f_0)-\Gamma(\bar B_s(t)\to J/\psi f_0)}{\Gamma(B_s(t)\to J/\psi f_0)
+\Gamma(\bar B_s(t)\to J/\psi f_0)}=\frac{C\cos(\Delta M_st) -
S \sin(\Delta M_st)}{\cosh(\Delta\Gamma_st/2)+
{\cal A}_{\Delta\Gamma}\sinh(\Delta\Gamma_st/2)} ,
\end{equation}
where $C$ and ${\cal A}_{\Delta\Gamma}$ are given in (\ref{ACPdir}) and (\ref{ADG}), respectively.
The ``mixing-induced" CP-violating observable
\begin{equation}\label{S-def}
S\equiv S(B_s\to J/\psi f_0) = \frac{- 2\,\mbox{Im}\,\xi_{J/\psi f_0}^{(s)}}{1+
\bigl|\xi_{J/\psi f_0}^{(s)}\bigr|^2}
\end{equation}
originates from interference between $B^0_s$--$\bar B^0_s$ mixing
and decay processes, and can be written with the help of $\Delta\phi$ introduced
in (\ref{sDelphi})--(\ref{tDelphi}) as follows \cite{FFM}:
\begin{equation}
S=\sqrt{1-C^2}\sin(\phi_s+\Delta\phi).
\end{equation}

\begin{figure}[t]
   \centering
   \includegraphics[width=7.9truecm]{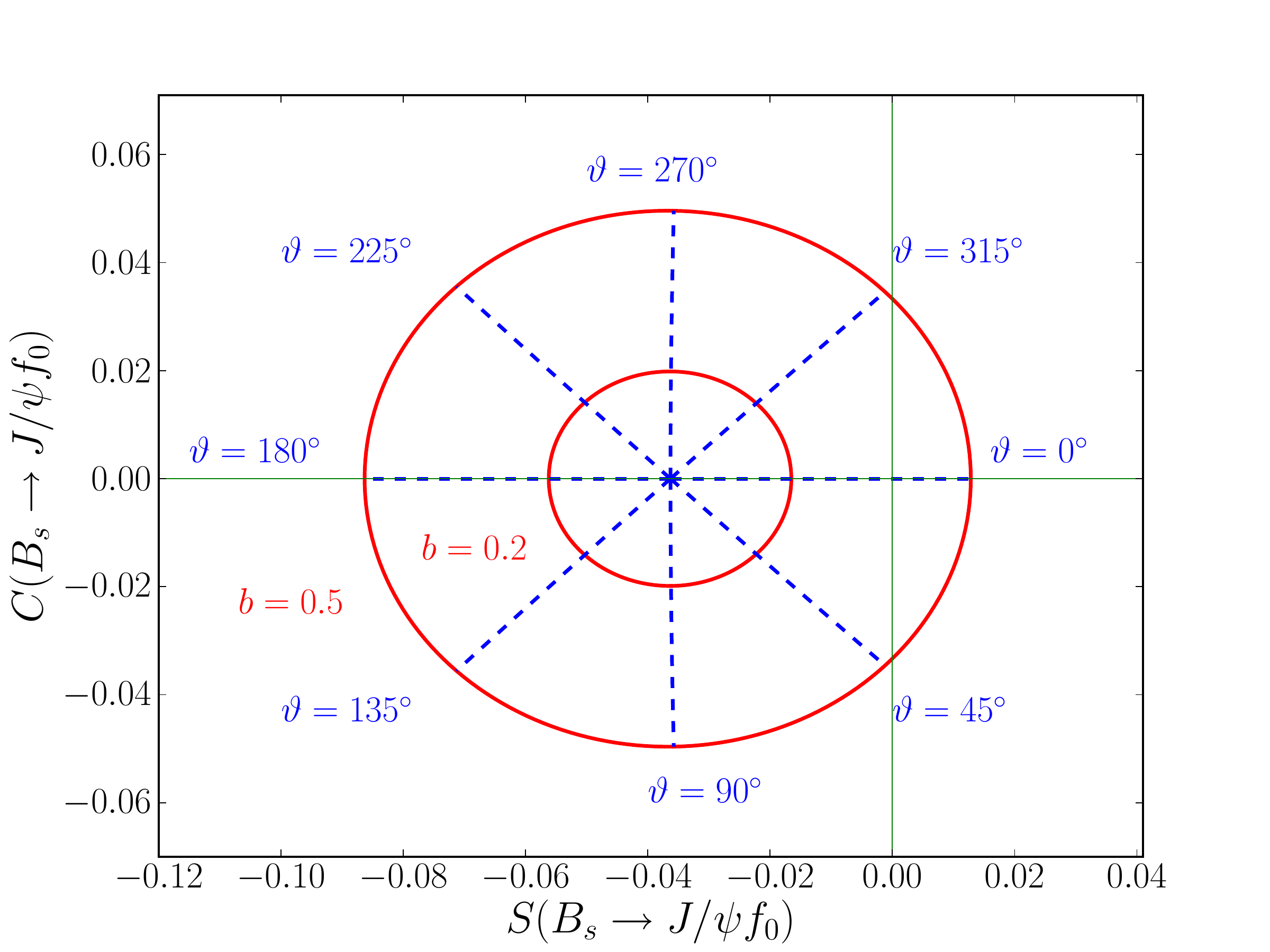}
   \caption{Parametric plot of the mixing-induced and direct CP asymmetries of the
   $B^0_s\to J/\psi f_0$ decay, $S$ and $C$, respectively, for $\gamma = 68^\circ$
   and the SM central value $\phi_s^{\rm SM} = -2.1^\circ$. The solid rings correspond to
   the fixed points of $b=0.2$ and 0.5 with $\vartheta$ allowed to vary. Likewise, the dashed
   lines are fixed points of $\vartheta$ with $b$ allowed to vary.}\label{fig:dirCP}
\end{figure}

In the left panel of Fig.~\ref{fig:CP}, we show the dependence of $S$ on the
$B^0_s$--$\bar B^0_s$ mixing phase $\phi_s$. In order to explore the impact of
the hadronic effects, we assume again (\ref{b-range}) with (\ref{gam-range}), and add the
resulting errors in quadrature to attain the band shown in the figure. Our SM prediction,
which is indicated by the error bar, is given by
\begin{equation}\label{S-SM}
\left.S(B_s^0 \to J/\psi f_0)\right|_{\rm SM} \in [ -0.086, -0.012],
\end{equation}
which should be compared with the na\"ive SM value, $(\sin\phi_s)|_{\rm SM}=-0.036\pm 0.002$,
corresponding to $b=0$. The right panel of Fig.~\ref{fig:CP} is a zoomed in version of the
same plot, focusing on smallish phases $\phi_s$. Here the individual errors associated with
the input parameters have been included, revealing that $b$ and $\vartheta$ lead to a
comparable  and sizable error in this $\phi_s$ domain, whereas the error on
$\gamma$ in (\ref{gam-range}) is negligible. These plots are complemented
by Fig.~\ref{fig:dirCP}, where we show the dependence of $S$ and $C$ on $(b, \vartheta)$
and the resulting correlation between these observables for the SM central value of $\phi_s$.

From these plots and the range in (\ref{S-SM}) we conclude that the measurement of the
mixing-induced CP violation in $B^0_s\to J/\psi f_0$ would give us unambiguous evidence
for NP in $B^0_s$--$\bar B^0_s$ mixing should a large value of $|S|$ be found. For instance,
$\phi_s\sim-45^\circ$ would correspond to $S\sim-70\%$. However, such a fortunate scenario
is now essentially excluded by the LHCb result given in (\ref{LHCb-phis}). Still, LHCb may
eventually measure sizeable mixing-induced CP violation in $B^0_s\to J/\psi f_0$. Should
its value fall into the range
\begin{equation}\label{S-limit}
-0.1\lsim S \lsim 0,
\end{equation}
the hadronic SM effects related to the $b$ parameter would preclude conclusions on
the presence or absence of CP-violating NP contributions to $B^0_s$--$\bar B^0_s$
mixing, unless we have insights into this parameter. First constraints can obviously be
obtained through the measurement of direct CP violation in $B^0_s\to J/\psi f_0$.
However, this asymmetry takes a value of at most $\sim 5\%$ in Fig.~\ref{fig:dirCP}
and will be challenging to measure precisely.

Before discussing a ``control" channel to constrain the $be^{i\theta}$ parameter through
data in the spirit of the strategy involving the $B^0_{s,d}\to J/\psi K_{\rm S}$ system
proposed in Ref.~\cite{RF-BpsiK}, let us spend a few words on the most recent developments.
The LHCb collaboration has reported the first (preliminary) result for the determination
of $\phi_s$ from the mixing-induced CP violation in the $B^0_s\to J/\psi f_0$ channel,
which reads as follows \cite{LHCb-f0-CP}:
\begin{equation}
\phi_s=-(25\pm25\pm1)^\circ,
\end{equation}
and corresponds to $S=-0.43^{+0.43}_{-0.34}$. In this analysis, the hadronic corrections
discussed above were not taken into account. There is still some way to go until we
may eventually enter the limiting range in (\ref{S-limit}) and it will be interesting to see
the evolution of the data.

The LHCb Collaboration has also obtained an average with the $B^0_s\to J/\psi \phi$ result
in (\ref{LHCb-phis}), which is given by $\phi_s=+(1.7\pm9.2\pm4.0)^\circ$ \cite{LHCb-f0-CP}.
Once the $B^0_s\to J/\psi f_0$ analysis becomes more precise, it will be problematic in view
of the hadronic effects and their different impact on the $B^0_s\to J/\psi f_0$ and
$B^0_s\to J/\psi \phi$ channels to make such an average. It will actually be very interesting
to compare the individual measurements as the precision increases, which may also
provide insights into the hadronic corrections.

\boldmath
\section{The $B^0_d\to J/\psi f_0(980)$ Channel}\label{sec:Bd}
\unboldmath
\setcounter{equation}{0}
\subsection{Decay Amplitude and Observables}
An interesting decay to obtain insights into the size of the hadronic parameter
$be^{i\vartheta}$ is $B^0_d\to J/\psi f_0(980)$, which we will abbreviate in the following
as $B^0_d\to J/\psi f_0$. In order to obtain its decay topologies, we have simply to
interchange all strange and down quarks in Fig.~\ref{fig:topol}.
The leading colour-suppressed tree-diagram-like topology emerges from
the $d\bar d$ component of the $f_0(980)$ state.

A key feature of the $B^0_d\to J/\psi f_0$ mode is that the CKM factors
$\lambda^{(d)}_q\equiv V_{qd} V_{qb}^*$ enter the expression for the decay amplitude.
If we assume the SM and apply the unitarity of the CKM matrix,  we arrive at
\begin{equation}\label{ampl-2}
A(B^0_d\to J/\psi f_0)=-\lambda{\cal A}' \left [1-  b' e^{i\vartheta'}e^{i\gamma}  \right],
\end{equation}
where ${\cal A}'$ and $b' e^{i\vartheta'}$ take the same form as \eqref{Acal} and \eqref{hadrDefn},
respectively. In contrast to the $B^0_s\to J/\psi f_0$ amplitude (\ref{ampl-1}), the
hadronic parameter $b'e^{i\vartheta'}$ is {\it not} suppressed by $\epsilon$. Consequently, its
impact is ``magnified" in $B^0_d\to J/\psi f_0$ with respect to $B^0_s\to J/\psi f_0$.

The CP-violating rate asymmetry of $B^0_d\to J/\psi f_0$ takes the form
\begin{equation}
\frac{\Gamma(B_d(t)\to J/\psi f_0)-\Gamma(\bar B_d(t)\to J/\psi f_0)}{\Gamma(B_d(t)\to J/\psi f_0)
+\Gamma(\bar B_d(t)\to J/\psi f_0)}=C'\cos(\Delta M_dt) -  S' \sin(\Delta M_dt),
\end{equation}
where we have taken into account that the width difference of the $B_d$-meson
system is negligibly small, in contrast to the $B_s$-meson system.
In analogy to (\ref{t-dep-asym}), the observable
\begin{equation}
C'\equiv C(B_d\to J/\psi f_0)=\frac{2  b'\sin\vartheta' \sin\gamma}{N'}
\end{equation}
with
\begin{equation}
N'\equiv1-2 b'\cos\vartheta'\cos\gamma+b'^2
\end{equation}
describes direct CP violation. On the other hand,
\begin{equation}
S'\equiv S(B_d\to J/\psi f_0)=\sqrt{1-C'^2}\sin(\phi_d+\Delta\phi'),
\end{equation}
where $\phi_d$ is the $B^0_d$--$\bar B^0_d$ mixing phase,\footnote{Taking the
corrections from doubly Cabibbo-suppressed penguin topologies through $B^0_d\to J/\psi \pi^0$ 
data into account, the mixing-induced CP violation in $B^0_d\to J/\psi K_{\rm S,L}$ gives 
$\phi_d=(42.2^{+3.4}_{-1.7})^\circ$ \cite{FFJM}.} describes mixing-induced CP violation.
The phase shift $\Delta\phi'$ can be obtained from
\begin{equation}
\sin\Delta\phi' = \frac{-2 b'\cos\vartheta' \sin\gamma+b'^2
\sin2\gamma}{N'\sqrt{1-C'^2}}
\end{equation}
\begin{equation}
\cos\Delta\phi'=\frac{1- 2 b'\cos\vartheta' \cos\gamma+b'^2
\cos2\gamma}{N'\sqrt{1-C'^2}},
\end{equation}
yielding
\begin{equation}
\tan\Delta\phi'=\frac{-2 b'\cos\vartheta' \sin\gamma+b'^2
\sin2\gamma}{1- 2 b'\cos\vartheta'\cos\gamma+b'^2\cos2\gamma}.
\end{equation}
The measurement of $C'$ and $S'$ allows us to determine $b'$ and $\theta'$; the corresponding
expressions can be obtained straightforwardly from the formulae given in Ref.~\cite{FFJM}.

\boldmath
\subsection{Specific Assumptions about the $f_0(980)$}
\unboldmath
As in Section~\ref{sec:assumptions}, let us now discuss the forms of the hadronic parameter
$b'e^{i\theta'}$ in the quark--antiquark and tetraquark frameworks. In the former case, in analogy to
the flavour decomposition for the $B_s^0\to J/\psi f_0$ channel in \eqref{BsFlavDecomp}, we
obtain for the $B_d\to J/\psi f_0$ decay:
\begin{equation}\label{BdDecomp}
A^{\prime(c)}_{\rm T} = \frac{\sin\mixAng}{\sqrt{2}} \tilde{A}^{\prime(c)}_{{\rm T},d\bar{d}},
\quad
A^{\prime(qt)}_{\rm P} = \frac{\sin\mixAng}{\sqrt{2}} \tilde{A}^{\prime(qt)}_{{\rm P},d\bar{d}},
\end{equation}
whereas the amplitudes $A^{\prime(c)}_{\rm E}$, $A^{\prime(u)}_{\rm E}$ and
$A^{\prime(qt)}_{\rm PA}$ take the same forms as their $B_s^0\to J/\psi f_0$
counterparts. Using $SU(3)_{\rm F}$ flavour symmetry, we can drop the $q\bar q$ subscripts.
Moreover, we can then also identify the $B_d^0\to J/\psi f_0$
amplitudes with their $B_s^0\to J/\psi f_0$ partners, i.e.\ can simply drop the primes.
This results in the following expression for the hadronic $B_d^0 \to J/\psi f_0$
parameter:
\begin{equation}
\left. b' e^{i\vartheta'}\right|_{q\bar{q}} = R_b\left[
	\frac{\cos\mixAng\left\{ \tilde{A}^{(ut)}_{\rm PA}\right\} +
	\frac{1}{\sqrt{2}}\sin\mixAng \left\{\tilde{A}^{(ut)}_{\rm P} + \tilde{A}^{(u)}_{\rm E} +
	2 \tilde{A}^{(ut)}_{\rm PA}\right\}}
	{\cos\mixAng\left\{ \tilde{A}^{(c)}_{\rm E} + \tilde{A}^{(ct)}_{\rm PA}\right\} +
	\frac{1}{\sqrt{2}}\sin\mixAng \left\{ \tilde{A}^{(c)}_{\rm T} + \tilde{A}^{(ct)}_{\rm P} + 2 \tilde{A}^
	{(c)}_{\rm E} +
	2 \tilde{A}^{(ct)}_{\rm PA}\right\}}\right],
	\label{bthetaPrime}
\end{equation}
which should be compared with  \eqref{btheta}.

In contrast to the conventional $SU(3)_{\rm F}$ strategies involving decays of $B_{(s)}$ mesons
into pions and kaons, there is a complication due to the hadronic structure of the $f_0(980)$,
which is reflected in (\ref{btheta}) and (\ref{bthetaPrime}) by the dependence on the mixing
angle $\mixAng$. There is an interesting situation, corresponding to
\begin{equation}\label{IM}
\cos\mixAng=\frac{1}{\sqrt{2}}\sin\mixAng,
\end{equation}
which is satisfied for $\mixAng=55^\circ$. In this case, we have
\begin{equation}
|f_0(980)\rangle=\frac{1}{\sqrt{3}}\left[ |u\bar u\rangle + |d\bar d\rangle + |s\bar s\rangle \right],
\end{equation}
and arrive at the following expression:
\begin{equation}\label{beqbp}
\left. b e^{i\vartheta} \right|_{q\bar{q}} =
R_b\left[
	\frac{\tilde{A}^{(ut)}_{\rm P}  +
	\tilde{A}^{(u)}_{\rm E} + 3 \tilde{A}^{(ut)}_{\rm PA} }
	{\tilde{A}^{(c)}_{\rm T} + \tilde{A}^{(ct)}_{\rm P} + 3 \tilde{A}^{(c)}_{\rm E} +
	3 \tilde{A}^{(ct)}_{\rm PA}}\right] = \left. b' e^{i\vartheta'} \right|_{q\bar{q}}.
\end{equation}
We could then simply identify the $be^{i\theta}$ of the $B^0_s\to J/\psi f_0$ channel
with the $b'e^{i\theta'}$ of the $B^0_d\to J/\psi f_0$ mode. Looking at the current ranges of
$\mixAng$ summarized in Section~\ref{sec:qqbar}, this scenario -- or a situation close to
it -- may actually be realized in nature. It is interesting to note that the flavour structure of
(\ref{IM}) corresponds to an $SU(3)_{\rm F}$ singlet, in analogy to the $\eta_1$ state
of the $\eta$--$\eta'$ system of the pseudo-scalar mesons.

On the contrary, as was discussed in Section \ref{sec:f0}, the tetraquark interpretation
of the $f_0(980)$ appears more favourable. In this picture,  we obtain
\begin{align}
A^{\prime(c)}_{\rm T} &= \frac{1}{\sqrt{2}} \tilde{A}^{\prime(c)}_{{\rm T},sd\bar{d}\bar{s}}\notag\\
A^{\prime(qt)}_{\rm P} &=  \frac{1}{\sqrt{2}}\tilde{A}^{\prime(qt)}_{{\rm P},sd\bar{d}\bar{s}}\notag\\
A^{\prime(c)}_{\rm E} &= \frac{1}{\sqrt{2}}\left[
	  \tilde{A}^{\prime(c)}_{{\rm E},su\bar{u}\bar{s}}
	+ \tilde{A}^{\prime(c)}_{{\rm E},sd\bar{d}\bar{s}}
	+ \tilde{A}^{\prime(c)}_{{\rm E},us\bar{s}\bar{u}}
	+ \tilde{A}^{\prime(c)}_{{\rm E},ds\bar{s}\bar{d}}
	\right]
	\stackrel{SU(3)_{\rm F}}{=} 2\sqrt{2} \tilde{A}^{\prime(c)}_{{\rm E}},\notag\\
A^{\prime(u)}_{\rm E} &=\frac{1}{\sqrt{2}} \tilde{A}^{\prime(u)}_{{\rm E},us\bar{s}\bar{u}}
	= \frac{1}{\sqrt{2}} \tilde{A}^{\prime(u)}_{{\rm E}},\notag\\
A^{\prime(qt)}_{\rm PA} &=  \frac{1}{\sqrt{2}}\left[
	  \tilde{A}^{\prime(c)}_{{\rm PA},su\bar{u}\bar{s}}
	+ \tilde{A}^{\prime(c)}_{{\rm PA},sd\bar{d}\bar{s}}
	+ \tilde{A}^{\prime(c)}_{{\rm PA},us\bar{s}\bar{u}}
	+ \tilde{A}^{\prime(c)}_{{\rm PA},ds\bar{s}\bar{d}}
	\right]
	\stackrel{SU(3)_{\rm F}}{=} 2\sqrt{2} \tilde{A}^{\prime(qt)}_{{\rm PA}},
\label{TetraDecompprime}
\end{align}
in analogy to (\ref{TetraDecomp}).
The $A_{4q}$ topology shown in Fig.~\ref{fig:f0topol} does not have a counterpart in
$B^0_d\to J/\psi f_0$ for $\omega=0$ in (\ref{tetra}), which was assumed in the
expressions given above. For a non-vanishing value of this angle, it would be
suppressed by $\sin\omega<0.1$. Assuming again the $SU(3)_{\rm F}$ flavour
symmetry to identify the topological amplitudes in $B^0_d\to J/\psi f_0$ and
$B^0_s\to J/\psi f_0$, we arrive at
\begin{equation}
\left. b' e^{i\vartheta'}\right|_{\rm 4q} = R_b\left[
	\frac{\tilde{A}^{(ut)}_{\rm P} +
	\tilde{A}^{(u)}_{\rm E} + 4 \tilde{A}^{(ut)}_{\rm PA}}
	{\tilde{A}^{(c)}_{\rm T} + \tilde{A}^{(ct)}_{\rm P} + 4\tilde{A}^{(c)}_{\rm E} + 4\tilde{A}^{(ct)}_{\rm
	PA}
	}\right].
	\label{bTetraprime}
\end{equation}
It is interesting to note that the $q\bar q$ expression (\ref{bthetaPrime}) reproduces the
form of the tetraquark expression (\ref{phi-4q}) for a mixing angle $\mixAng$ satisfying
(\ref{phi-4q}), although the individual topological amplitudes would in general take
different values in the $q\bar q$ and $4q$ frameworks.

\boldmath
\subsection{Estimate of the $B^0_d\to J/\psi f_0$ Branching Ratio}
\unboldmath
For experimental studies, it is useful to estimate the branching ratio of the
$B^0_d\to J/\psi f_0$ decay. Using (\ref{ampl-1}) and (\ref{ampl-2}), we obtain
the following expression for the ratio of the CP-averaged decay amplitudes:
\begin{equation}
\left|\frac{\langle A(B_d\to J/\psi f_0)\rangle}{\langle A(B_s\to J/\psi f_0)\rangle}\right|^2
=\epsilon\left[\frac{1-2b'\cos\vartheta'\cos\gamma+b'^2}{1+2\epsilon
b\cos\vartheta\cos\gamma+\epsilon^2b^2}\right]
\left|\frac{{\cal A}'}{{\cal A}}\right|^2,
\end{equation}
where (\ref{Acal}) gives
\begin{equation}
\left|\frac{{\cal A}'}{{\cal A}}\right|=
\left|\frac{A^{\prime(c)}_{\rm T}+A_{\rm P}^{\prime(ct)}+A_{\rm E}^{\prime(c)} +
A_{\rm PA}^{\prime(ct)}}{A^{(c)}_{\rm T}+
A_{\rm P}^{(ct)}+A_{\rm E}^{(c)} +A_{\rm PA}^{(ct)}}\right|.
\end{equation}
Keeping only the tree and penguin contributions and using the $SU(3)_{\rm F}$ symmetry yields
\begin{equation}\label{A-rat}
\left|\frac{{\cal A}'}{{\cal A}}\right|_{\rm q\bar q}=\frac{\tan\mixAng}{\sqrt{2}} \quad\mbox{and}\quad
\left|\frac{{\cal A}'}{{\cal A}}\right|_{4q}=\frac{1}{2}
\end{equation}
for the quark--antiquark and tetraquark descriptions of the $f_0(980)$, respectively.
For the former case, the result
\begin{equation}\label{Fd-FF}
	\left|\frac{ {\cal A}'}{\cal A}\right|_{q\bar{q}}\sim \left[\frac{F_1^{B^0_df_0}(M_{J/\psi}^2)}{F_1^
	{B^0_sf_0}(M_{J/\psi}^2)}\right]_{\mixAng=41.6^\circ}
\sim 0.44,
\end{equation}
which was obtained in Ref.~\cite{ElBennich:2008xy} for the $q\bar q$ framework
(using dispersion relations, see Section~\ref{sec:BRest}), is in the same ball-park as
\begin{equation}
	\left.\frac{\tan\mixAng}{\sqrt{2}}\right|_{\mixAng=41.6^\circ}=0.63.
\end{equation}

If we introduce the quantity
\begin{equation}\label{H-f0}
H_{f_0}\equiv\frac{1-2b'\cos\vartheta'\cos\gamma+b'^2}{1+2\epsilon
b\cos\vartheta\cos\gamma+\epsilon^2b^2},
\end{equation}
we can write the CP-averaged branching ratio as
\begin{equation}\label{eqn:brH}
\mbox{BR}(B^0_d\to J/\psi f_0)=H_{f_0} \times \mbox{BR}(B^0_d\to J/\psi f_0)_0,
\end{equation}
where
\begin{equation}\label{eqn:BrdZero}
\mbox{BR}(B^0_d\to J/\psi f_0)_0=\epsilon
\left|\frac{{\cal A}'}{{\cal A}}\right|^2
\left(\frac{M_{B^0_d}\Phi_d'}{M_{B^0_s}\Phi_s}\right)^3\frac{\tau_{B^0_d}}{\tau_{B^0_s}}\,
 \mbox{BR}(B^0_s\to J/\psi f_0)
\end{equation}
is the CP-averaged branching ratio in the limit $b'=0$; $\Phi_d'$ denotes the
$B^0_d\to J/\psi f_0$ phase-space factor.
This relation holds correspondingly for the branching ratios with
$f_0\to\pi^+\pi^-$. Using (\ref{eqn:Bf0PiPi}) yields
\begin{equation}\label{BR0-est}
{\rm BR}(B_d^0 \to J/\psi f_0; f_0 \to \pi^+ \pi^-)_0 \sim \left(1.65^{+0.34}_{-0.29} \right)
\times 10^{-6},
\end{equation}
where we have used the tetraquark value in (\ref{A-rat}), which is also in the ball-park
of (\ref{Fd-FF}). In this estimate, the error is essentially due to (\ref{eqn:Bf0PiPi}) and
does not take (unknown) theoretical uncertainties into account. As we will see in
Section~\ref{sec:hadr-contr}, the range $0\leq b' \lsim 0.5$ corresponds to
$0.8\lsim H_{f_0}\lsim 1.6$, so that (\ref{eqn:brH}) yields for the central value
in (\ref{BR0-est}) the following range:
\begin{equation} \label{BR-est}
\mbox{BR}(B^0_d\to J/\psi f_0; f_0 \to \pi^+ \pi^-)\sim (1\mbox{--}3)\times 10^{-6}.
\end{equation}

Since the tetraquark picture corresponds to (\ref{phi-4q}) with $\mixAng=35^\circ$,
it is more predictive for the estimate of the $B^0_d\to J/\psi f_0$ branching ratio than
the quark--antiquark framework. As we discussed in Section~\ref{sec:qqbar}, in the latter case,
the mixing angle suffers from large uncertainties.  Obviously, as the leading contribution
to $B_d^0 \to J/\psi f_0$ is caused by the $d\bar d$ component of the $f_0(980)$, a
mixing angle close to $0^\circ$ or $180^\circ$ would strongly suppress the decay.
On the other hand, the observation of $B_d^0 \to J/\psi f_0$  in the $10^{-6}$ regime
would imply a significant $d\bar d$ component of the $f_0(980)$. It is interesting to
note that about four times more $B_d$ than $B_s$ mesons are produced at the
Tevatron and LHCb \cite{HFAG,LHCb-fsfd}, which partly compensates the CKM
suppression of the $B^0_d\to J/\psi f_0$ channel with respect to $B^0_s\to J/\psi f_0$.
In view of (\ref{BR-est}), the first signals of the $B^0_d\to J/\psi f_0$ decay with
$f_0\to\pi^+\pi^-$ may be seen in the near future. Needless to note, the $B^0_d\to J/\psi f_0$
is also an interesting topic for the $e^+e^-$ SuperKEKB and SuperB projects.

\subsection{Hierarchy of Topological Amplitudes}\label{sec:hier}
So far, we have not assumed any hierarchy for the different topologies contributing to the
decays at hand. The dominant contribution is expected to be given by the colour-suppressed
tree amplitude $A_{\rm T}^{(c)}$. Should all other topologies give negligible contributions,
we would simply have $b'=b=0$, and the observables discussed in Sections~\ref{sec:lifetime}
and \ref{sec:CP} would not be affected by hadronic uncertainties and the structure of the
$f_0(980)$.

Should in addition to $A_{\rm T}^{(c)}$ only the penguin topologies described by the
$A_{\rm P}^{(q)}$ amplitudes have a significant impact, thereby resulting in a sizable value
of $b$, the situation would be given in the
$SU(3)_{\rm F}$ limit as follows:
\begin{equation}\label{SU3-rel}
be^{i\vartheta}=R_b\left[\frac{A_{\rm P}^{(ut)}}{A_{\rm T}^{(c)}+A_{\rm P}^{(ct)}}\right]=
b'e^{i\vartheta'},
\end{equation}
both in the tetraquark and $q\bar q$ descriptions of the $f_0(980)$. In the latter case, however,
we have to assume that the mixing angle $\mixAng$ is significantly different
from $0^\circ$ or $180^\circ$, as is evident from (\ref{BdDecomp}) and (\ref{bthetaPrime});
the ideal situation would correspond to (\ref{IM}), yielding (\ref{beqbp}). The hadronic corrections
to the mixing-induced CP violation in $B^0_s\to J/\psi f_0$ could then be constrained through
the $B^0_d\to J/\psi f_0$ mode.

In addition to $SU(3)_{\rm F}$-breaking effects, the relation in (\ref{SU3-rel}) is affected by the
additional topologies. The exchange and penguin annihilation topologies, which involve the
spectator quarks, are usually neglected in the literature (see, for instance,
Refs.~\cite{DGR,skands,CGR}). In the case of $B$ decays involving the $f_0(980)$, there is an
interesting argument, which supports their suppression, that is related to the decay constant of
this state. Namely, the $f_0(980)$ decay constant is defined by
\begin{equation}\label{f-def}
\langle f_0(p)| \bar q \gamma^{\mu}(1-\gamma_5)q |0\rangle=
\langle f_0(p)| \bar q \gamma^{\mu} q |0\rangle \equiv f_{f_0}p^\mu,
\end{equation}
where the axial-vector current does not contribute because of Lorentz symmetry
($q\in\{s,d,u\}$).
Using the CP transformation
\begin{equation}
({\cal CP})\left[ \bar q \gamma^{\mu}(1-\gamma_5)q\right] ({\cal CP})^\dagger = -
\left[ \bar q \gamma_{\mu}(1-\gamma_5)q\right]
\end{equation}
with $({\cal CP})^\dagger ({\cal CP})=\hat 1$ and $({\cal CP})|f_0\rangle=+|f_0\rangle$
in (\ref{f-def}) as well as the CP invariance of strong interactions, it follows straightforwardly that
the decay constant has to vanish, i.e.\ $f_{f_0}=0$. The same argument in fact applies to all
CP-selfconjugate scalar states. Consequently, the exchange and penguin annihilation
topologies will vanish in the factorization picture as, in this framework, they are proportional to the
product $f_{B_{s,d}}\, f_{J/\psi}\, f_{f_0}$ of the decay constants.
This feature suggests that these topologies play an even less pronounced
role than they do in $B$ decays into pseudo-scalar/vector mesons.
Moreover, it is plausible to
assume that they are suppressed significantly with respect to the penguin contributions
$A_{\rm P}^{(qt)}$ of the $B^0_{s,d}\to J/\psi f_0$ decays.

\begin{figure}[tbp] 
   \centering
   \includegraphics[width=15cm]{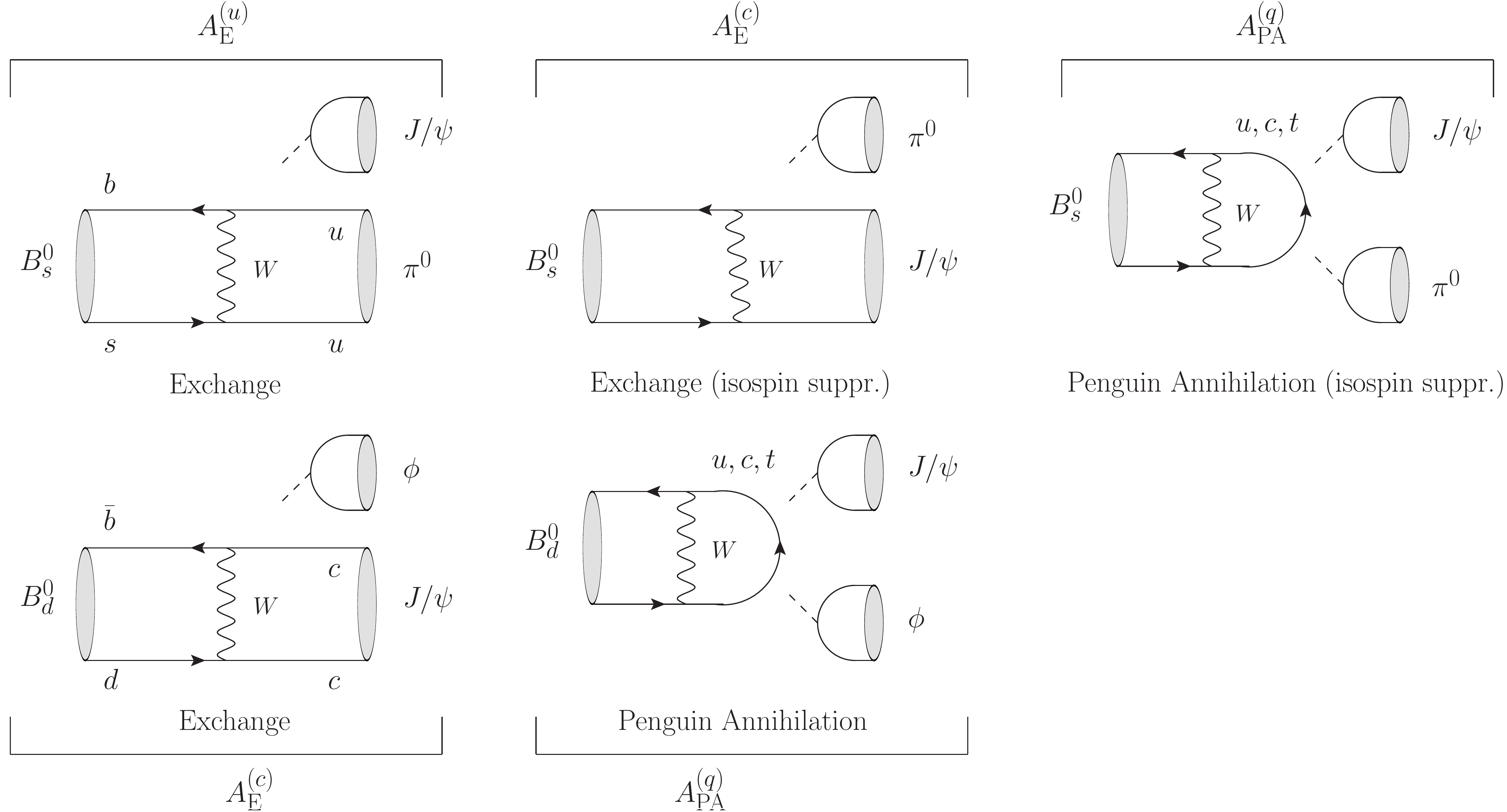}
   \caption{Decay topologies contributing to the $B^0_s\to J/\psi \pi^0$ (top row)
   and $B^0_d\to J/\psi \phi$ (bottom row) decays.
   The creation of a $\pi^0$ from a colourless state is forbidden by isospin symmetry
   (QED effects and EW penguin annihilation topologies can circumvent this
   argument).}\label{fig:control}
\end{figure}

Experimental insights into this issue for $B$ decays into ``conventional" mesons can
be obtained through the $B^0_d\to J/\psi \phi$ \cite{FFM} and $B^0_s\to J/\psi \pi^0$
modes \cite{FFJM}, which can only emerge from exchange and penguin annihilation
topologies. As can be seen in Fig.~\ref{fig:control}, the $B^0_d\to J/\psi \phi$ and
$B^0_s\to J/\psi \pi^0$ decays probe the counterparts of the $A_{\rm E}^{(c)}+A_{\rm PA}^{(ct)}$
and $A_{\rm E}^{(u)}+A_{\rm PA}^{(ct)}$ amplitudes, respectively. The current experimental
upper bounds on the branching ratios can be summarized as follows (90\% C.L.):
\begin{eqnarray}
\mbox{BR}(B^0_d\to J/\psi \phi)&<& 9.4\times10^{-7} \, \mbox{\cite{Belle-psiphi},} \label{Bd-constr}\\
\mbox{BR}(B^0_s\to J/\psi \pi^0)&<&1.2\times10^{-3} \, \mbox{\cite{L3}.} \label{Bs-constr}
\end{eqnarray}
If we use $\mbox{BR}(B^0_d\to J/\psi K^{*0})=(1.33\pm0.06)\times10^{-3}$ \cite{PDG}
and the $SU(3)$ flavour symmetry, the upper bound in (\ref{Bd-constr}) allows us to
obtain the following constraint:
\begin{equation}\label{b-1}
\left|\frac{A_{\rm E}^{(c)}+A_{\rm PA}^{(ct)}}{A_{\rm T}^{(c)}}
\right|\sim\left(\frac{1-\lambda^2/2}{\lambda}\right)
\sqrt{\frac{\mbox{BR}(B^0_d\to J/\psi \phi)}{\mbox{BR}(B^0_d\to J/\psi K^{*0})}}
\lsim 0.1.
\end{equation}
Here $A_{\rm E}^{(c)}$, $A_{\rm PA}^{(ct)}$ and $A_{\rm T}^{(c)}$
denote exchange, penguin annihilation and colour-suppressed tree amplitudes in
these decays, which are the counterparts of those contributing to $B^0_{s,d}\to J/\psi f_0$.
Since we have two vector mesons in the final state, angular distributions should be used
to disentangle the different final-state configurations. For simplicity, we have just assumed
``generic" sizes for the topological amplitudes. The upper bound in (\ref{b-1}) supports
the expectation that the exchange and penguin annihilation topologies are strongly suppressed.
It would be important to further improve the upper bound in (\ref{Bd-constr}) and to put constraints
on the $B^0_s\to J/\psi \pi^0$ branching ratio that are much more stringent than the
one in (\ref{Bs-constr}), which was published by L3 in 1997.

The ``scalar-meson" counterpart of $B^0_s\to J/\psi \pi^0$  is given by the
$B^0_s\to J/\psi a_0^0(980)$ channel, where
\begin{equation}
a^0_0(980)=\frac{1}{\sqrt{2}}\left(u\bar u - d\bar d\right) \quad\mbox{and}\quad
a^0_0(980)=\frac{1}{\sqrt{2}}\left([su][\bar s \bar u]-[sd][\bar s\bar d]\right)
\end{equation}
in the quark--antiquark and tetraquark pictures, respectively.
If we neglect the isospin-suppressed topologies corresponding to those in Fig.~\ref{fig:control},
we only get a contribution from the exchange topology in the quark--antiquark description
of the $a_0^0(980)$. On the other hand, in the tetraquark picture, we get an additional
contribution from the counterpart of the $A_{4q}$ topology in Fig.~\ref{fig:f0topol}.
Upper bounds on the branching ratio of the $B^0_s\to J/\psi a_0^0$ channel and their
comparison with $B^0_s\to J/\psi \pi^0$ would therefore allow us to put some constraints
on the $A_{4q}$ contribution.

Another interesting decay in this context is $B^0_s\to J/\psi \bar \kappa^0(800)$,
which receives only contributions from colour-suppressed tree and penguin topologies in
the quark--antiquark picture of the scalar state $\bar\kappa^0(800)=s\bar d$. On the other hand,
in the tetraquark description, $\bar \kappa^0=[su][\bar u\bar d]$, we get an additional
contribution from the counterpart of the $A_{4q}$ topology. However, the properties of
the $\kappa$ meson, which appears to have a very large width around 500~MeV
and sits close to the $K\pi$ threshold, are essentially unknown \cite{PDG}.

\begin{figure}[t]
   \centering
   \begin{tabular}{cc}
   	  \includegraphics[width=7.9truecm]{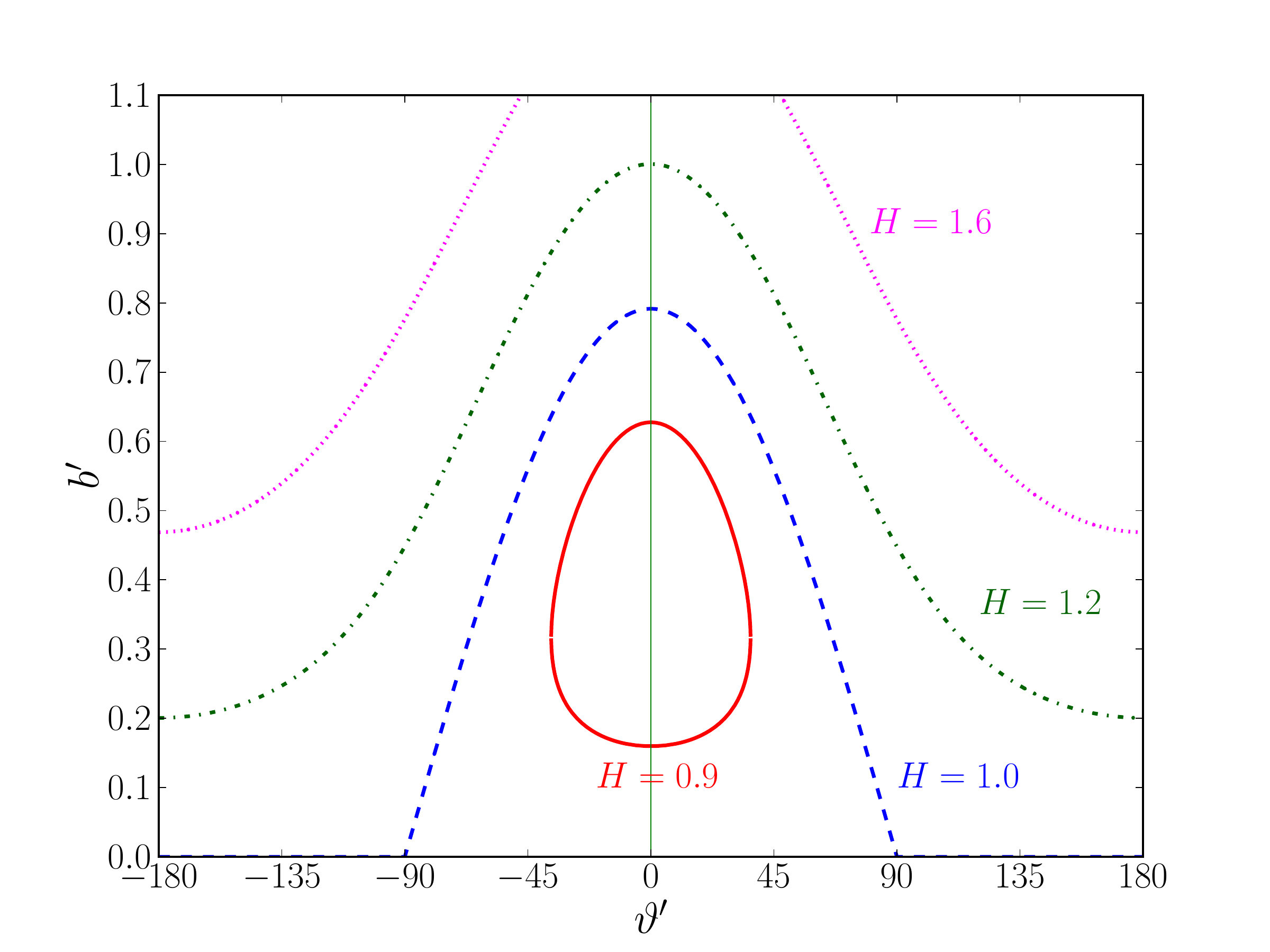} &
   	  \includegraphics[width=7.9truecm]{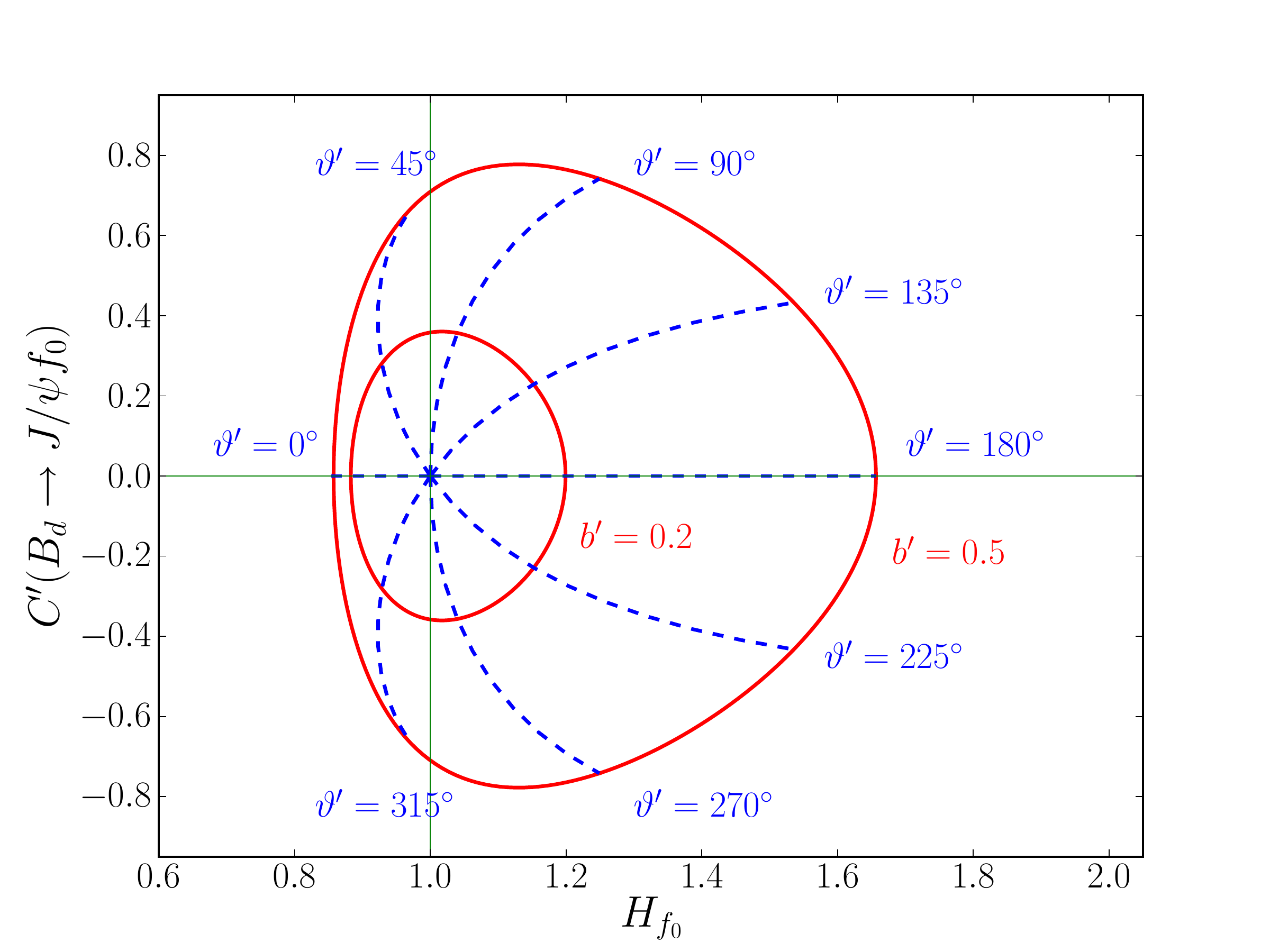}
   \end{tabular}
   \caption{ {\it Left panel:} Correlation between $b'$ and $\vartheta'$ for $H_{f_0}=0.9$, 1.0, 1.2 
   and 1.6.  {\it Right panel:} Correlation between $H_{f_0}$ and the direct CP asymmetry $C'$ of
   $B^0_d\to J/\psi f_0$, assuming $\gamma = 68^\circ$. The solid rings correspond
   to $b'=0.2$ and 0.5 with $\vartheta'$ allowed to vary. Likewise, the
   dashed lines are fixed points of $\vartheta'$ with $b'$ allowed to vary.}\label{fig:H}
\end{figure}

\boldmath
\subsection{Control of the Hadronic Effects in $B^0_s\to J/\psi f_0$}\label{sec:hadr-contr}
\unboldmath
Once the CP-averaged branching ratio of $B^0_d\to J/\psi f_0$ has been measured,
we can determine $H_{f_0}$ introduced in (\ref{H-f0}) by rewriting expressions \eqref{eqn:brH}
and \eqref{eqn:BrdZero} as follows:
\begin{equation}
	H_{f_0} = \frac{1}{\epsilon}
	\left|\frac{{\cal A}}{{\cal A}'}\right|^2
	\left(\frac{M_{B^0_s}\Phi_s}{M_{B^0_d}\Phi_d'}\right)^3\frac{\tau_{B^0_s}}{\tau_{B^0_d}}\,
	\frac{\mbox{BR}(B^0_d\to J/\psi f_0)}{\mbox{BR}(B^0_s\to J/\psi f_0)}.
\end{equation}
In the left panel of Fig.~\ref{fig:H}, we show the correlation between the hadronic parameters for
various values of $H_{f_0}$, assuming $b=b'$ and $\vartheta=\vartheta'$ in \eqref{H-f0}; even
dramatic corrections to these relations would have a tiny impact because of the
$\epsilon$ suppression in (\ref{H-f0}). Under the same assumption, we also show in the right
panel of Fig.~\ref{fig:H} the correlation between $H_{f_0}$ and the direct CP asymmetry $C'$ of
the $B^0_d\to J/\psi f_0$ channel.  First constraints on $b'$ can be obtained from $H_{f_0}$ as
follows:
\begin{equation}
(b')^{\rm max}_{\rm min}=\left|\left(\frac{1+\epsilon H_{f_0}}{1-\epsilon^2H_{f_0}}\right)\cos\gamma
\pm\sqrt{\left[\left(\frac{1+\epsilon H_{f_0}}{1-\epsilon^2 H_{f_0}}\right)\cos\gamma
\right]^2+\frac{H_{f_0}-1}{1-\epsilon^2H_{f_0}}} \right|,
\end{equation}
which corresponds for $\epsilon=0$ to the bounds derived in Ref.~\cite{RF-BpiK},
and could be significantly sharpened through the measurement of $C'$.
The major uncertainty of $H_{f_0}$ enters through the $SU(3)_{\rm F}$-breaking
amplitude ratio $|{\cal A}/{\cal A}'|$. In addition, if the mixing-induced CP asymmetry
$S'$ of $B^0_d\to J/\psi f_0$ is measured, the quantity $H_{f_0}$ would no longer be
needed for the determination of $b'$ and $\theta'$. It would then actually be interesting
to extract $|{\cal A}/{\cal A}'|$ from the data and to compare with (\ref{A-rat}) or better estimates 
of this ratio that may then be available. 

As we have seen in the previous section, we expect the exchange and penguin annihilation
topologies to play a minor role in the $B^0_{s,d}\to J/\psi f_0$ decays. In the quark--antiquark
picture, assuming that the mixing angle $\mixAng$ is significantly different
from $0^\circ$ or $180^\circ$, we would then have (\ref{SU3-rel}) in the $SU(3)_{\rm F}$
limit, and could control the hadronic effects in the mixing-induced
CP asymmetry of the $B^0_s\to J/\psi f_0$ decay.\footnote{In the case of $\mixAng$ close
to $0^\circ$ or $180^\circ$, the $B^0_d\to J/\psi a_0^0(980)$ channel offers an
alternative to $B^0_d\to J/\psi f_0$. It is the ``scalar-meson" counterpart of the
$B^0_d\to J/\psi \pi^0$ decay \cite{CPS,FFJM}.} The theoretical uncertainties are governed
by $SU(3)_{\rm F}$-breaking corrections and the situation would be similar to
$B^0_s\to J/\psi \phi$, as discussed in Ref.~\cite{FFM}. However, in contrast to this decay,
we also have to assume in the $B^0_s\to J/\psi f_0$ case that the $f_0(980)$ is a
quark--antiquark state with wave function (\ref{f0-structure}), which is far from being
established.

As we have seen in Section~\ref{sec:f0}, the tetraquark description of the $f_0(980)$
has a variety of phenomenological advantages. But in this framework we have
to deal with the additional topology $A_{4q}$ shown in Fig.~\ref{fig:f0topol}, which contributes
to $B^0_s\to J/\psi f_0$ but does not have a counterpart in $B^0_d\to J/\psi f_0$.
Can we make quantitative statements about the $A_{4q}$ topology? As can be seen in
Fig.~\ref{fig:f0topol}, it involves the production of  the $J/\psi$ through a colour-singlet
exchange, in analogy to the penguin and exchange topologies in Fig.~\ref{fig:topol}.
However, this topology contributes also in the ``spectator" approximation and arises at the
tree level, i.e.\ is not loop-suppressed like the penguin contributions. On the other hand,
the $us$ diquark and the $\bar u\bar s$  anti-diquark have to be produced in the decay
of the $b$ quark in such a way as to form the $f_0(980)$ bound state, which suggests a possible suppression.
Presumably strong attractive forces are at work between these quark
correlations, but the hadronization mechanism itself is essentially unknown.

The central question for the analysis of the $B^0_{s,d}\to J/\psi f_0$ system is the competition
between the $4q$ and penguin topologies in (\ref{bTetra}). Should the former give a contribution
at the same -- or even larger -- level as the penguins, which would be reflected by a sizeable
value of $b$, we could not control the hadronic effects through the $B^0_d\to J/\psi f_0$
channel. In view of this situation, more detailed studies of the $b$ parameter
in the tetraquark description of the $f_0(980)$ would be very important.

\section{Conclusions}\label{sec:concl}
\setcounter{equation}{0}
Thanks to recent measurements, the $B^0_s\to J/\psi f_0$ channel is receiving increasing
interest to complement the $B^0_s\to J/\psi \phi$ mode in the search for CP-violating NP
contributions to $B^0_s$--$\bar B^0_s$ mixing. In contrast to the latter decay, the
$B^0_s\to J/\psi f_0$ analysis is simpler as no time-dependent angular analysis is
required. On the other hand, we have approximately one fourth of the events available.

In this paper, we have performed a detailed study of the effective lifetime
and the CP-violating observables of $B^0_s\to J/\psi f_0$, with a critical look at possible
hadronic uncertainties. This is an important issue, in particular as the hadronic structure
of the $f_0(980)$ is still controversial. It turns out that the effective lifetime is very robust with
respect to such effects, with an error that is essentially fully dominated by the theoretical prediction
of the width difference of the $B_s$-meson system in the SM. We find that the first measurement
by the CDF collaboration is about $1\,\sigma$ above a general upper bound derived in this
paper, which relies on the SM value of $\Delta\Gamma_s/\Gamma_s$. A future measurement
of the lifetime at the 1\% level would offer an interesting probe of the $B^0_s$--$\bar B^0_s$
mixing phase, which may be affected by CP-violating contributions to $B^0_s$--$\bar B^0_s$
mixing.

A sharper picture of such effects is offered by the mixing-induced CP violation $S$ of the
$B^0_s\to J/\psi f_0$ channel. The LHCb collaboration has very recently reported the
first analysis of this kind, corresponding to $S=-0.42^{+0.42}_{-0.34}$.
Should the measured value of $S$ eventually fall into the range $-0.1\lsim S \lsim 0$,
hadronic SM effects would preclude us from drawing conclusions on the presence or
absence of CP-violating NP contributions to $B^0_s$--$\bar B^0_s$ mixing, unless
we have insights into these corrections.

The $B^0_d\to J/\psi f_0$ decay, which has not yet been observed, offers an interesting
control channel for the SM corrections. The leading contributions emerge from the $d\bar d$
component of the $f_0(980)$. It would be interesting to add $B^0_d\to J/\psi f_0$ to the
experimental agenda. We have estimated its branching ratio with $f_0\to\pi^+\pi^-$
at the few times $10^{-6}$ level. In this case, a first signal may be seen in the near future.

In the quark--antiquark description of the $f_0(980)$, assuming
a mixing angle $\mixAng$ significantly different from $0^\circ$ or $180^\circ$, we have
shown that the hadronic corrections to $S$ can be controlled through the observables
of $B^0_d\to J/\psi f_0$  by means of $SU(3)_{\rm F}$ arguments.
On the other hand, should the $f_0(980)$ be a tetraquark, we would have to deal with
an additional topology at the tree level in $B^0_s\to J/\psi f_0$ that does not have a counterpart
in $B^0_d\to J/\psi f_0$. If it plays a significant role with respect to the conventional
hadronic tree amplitude, there may be significant corrections for $S$ which can no longer be
controlled via $B^0_d\to J/\psi f_0$. 
In the event that future ``na\"ive"
determinations (i.e.\ neglecting the hadronic corrections) of $\phi_s$ from
$B^0_s\to J/\psi f_0$ and $B^0_s\to J/\psi \phi$ give sizeably different values,
this may be traced back to the tetraquark topology in $B^0_s\to J/\psi f_0$.

As the experimental precision improves, the SM corrections to the determination
of $\phi_s$ from the $B^0_s\to J/\psi \phi$ decay also have to be controlled. The conceptual
advantage of this channel with respect to $B^0_s\to J/\psi f_0$ is that it does not suffer
from the poorly known hadronic structure of the $f_0(980)$ state. Interesting control channels
for $B^0_s\to J/\psi \phi$ are $B^0_s\to J/\psi \bar K^{*0}$ and $B^0_d\to J/\psi\rho^0$.

We very much look forward to future theoretical insights into the structure of the $f_0(980)$
as well as to future measurements!

\vspace*{0.5truecm}

\noindent
{\it Acknowledgements}\\
We would like to thank Andrzej Buras for useful discussions.
G.R. is grateful to Nikhef, and in particular the theory group, for the hospitality during part of this
work  and for providing a friendly welcoming atmosphere.

\end{document}